\documentclass[preprint2]{aastex}

\newcommand{\arcyear}{\mbox{\ensuremath{^{\prime\prime}} yr\ensuremath{^{-1}}}}
\newcommand{\kmsec}{\mbox{km sec\ensuremath{^{-1}}}}
\newcommand{\lsim}{\mathrel{\rlap{\lower4.5pt\hbox{$\sim$}}\raise2pt\hbox{$<$}}}
\newcommand{\msun}{$M_\odot$}
\newcommand{\lsun}{$L_\odot$}
\newcommand{\logg}{log $g$ }
\newcommand{\vtan}{$V_{\rm tan}$ }

\usepackage{url}
\usepackage{graphicx}

\shorttitle{Solar Neighborhood. XXXIX.}
\shortauthors{Subasavage et al.}

\begin{document}


\title{The Solar Neighborhood. XXXIX. Parallax Results from the CTIOPI
and NOFS Programs: 50 New Members of the 25 Parsec White Dwarf Sample}


\author{John P.~Subasavage\altaffilmark{1,8}, Wei-Chun
  Jao\altaffilmark{2,8}, Todd J.~Henry\altaffilmark{3,8}, Hugh
  C.~Harris\altaffilmark{1}, Conard C.~Dahn\altaffilmark{1},
  P.~Bergeron\altaffilmark{4}, P.~Dufour\altaffilmark{4}, Bart
  H.~Dunlap\altaffilmark{5}, Brad N.~Barlow\altaffilmark{6}, Philip
  A.\ Ianna\altaffilmark{3,8}, S{\'e}bastien
  L{\'e}pine\altaffilmark{2}, Steven J.~Margheim\altaffilmark{7}}











\altaffiltext{1}{U.S. Naval Observatory, 10391 West Naval Observatory Road, Flagstaff, 
AZ 86005-8521, USA; jsubasavage@nofs.navy.mil}
\altaffiltext{2}{Department of Physics and Astronomy, Georgia State University,
Atlanta, GA, 30302-4106, USA}
\altaffiltext{3}{RECONS Institute, Chambersburg, PA 17201, USA}
\altaffiltext{4}{D\'{e}partement de Physique, Universit\'{e} de Montr\'{e}al,
C.P. 6128, Succ. Centre-Ville, Montr\'{e}al, Qu\'{e}bec H3C 3J7,
Canada}
\altaffiltext{5}{University of North Carolina at Chapel Hill, Dept. of Physics
  and Astronomy, Chapel Hill, NC, 27599, USA}
\altaffiltext{6}{Department of Physics, High Point University, One
  University Parkway, High Point, NC, 27268, USA}
\altaffiltext{7}{Gemini Observatory, Southern Operations Center, Casilla 603, La
  Serena, Chile}
\altaffiltext{8}{Visiting astronomer, Cerro Tololo Inter-American
Observatory, National Optical Astronomy Observatory, which are
operated by the Association of Universities for Research in Astronomy,
under contract with the National Science Foundation.}


\begin{abstract}

We present 114 trigonometric parallaxes for 107 nearby white dwarf
(WD) systems from both the Cerro Tololo Inter-American Parallax
Investigation (CTIOPI) and the U.~S.~Naval Observatory Flagstaff
Station (NOFS) parallax programs.  Of these, 76 parallaxes for 69
systems were measured by the CTIOPI program and 38 parallaxes for as
many systems were measured by the NOFS program.  A total of 50 systems
are confirmed to be within the 25 pc horizon of interest.  Coupled
with a spectroscopic confirmation of a common proper motion companion
to a {\it Hipparcos} star within 25 pc as well as confirmation
parallax determinations for two WD systems included in the recently
released {\it Tycho Gaia} Astrometric Solution (TGAS) catalog, we add
53 new systems to the 25 pc WD sample --- a 42\% increase.  Our sample
presented here includes four strong candidate halo systems, a new
metal-rich DAZ WD, a confirmation of a recently discovered nearby
short-period (P = 2.85 hr) double degenerate, a WD with a new
astrometric perturbation (long period, unconstrained with our data),
and a new triple system where the WD companion main-sequence star has
an astrometric perturbation (P $\sim$ 1.6 yr).

\end{abstract}


\keywords{astrometry --- Galaxy: evolution --- solar neighborhood ---
stars: distances --- white dwarfs}



\section{Introduction}

White dwarfs (WDs) are the remarkably abundant remnants of the vast
majority of stars and serve as reliable tracers for a number of
astrophysically interesting topics.  We aim to compile a robust,
volume-limited sample of WDs upon which statistical studies can be
performed.  For instance, insight into population membership
percentages (thin disk, thick disk, halo), population ages, and
Galactic star formation history can be ascertained from the sample as
a whole.  Individually, a non-trivial number of WDs are found to be
metal-enriched and are most likely displaying signs of disrupted
planetary systems \citep[][and reference
  therein]{2009ApJ...694..805F}.  Given that the nearest
metal-enriched WDs are the brightest examples, they can be more
carefully studied.  With a volume-limited sample, one minimizes biases
in the WD luminosity function, mass function, and those introduced
when inferring the number of planetary systems around WDs.

We present here a compilation of two long-term astrometric efforts to
measure accurate trigonometric parallaxes to nearby WDs --- the Cerro
Tololo Inter-American Observatory Parallax Investigation
\citep[CTIOPI,][]{2005AJ....129.1954J} and the U.S.~Naval Observatory
Flagstaff Station \citep[NOFS,][]{1992AJ....103..638M} parallax
program.  In total, 76 parallaxes are measured for 69 systems by the
CTIOPI program and 38 parallaxes are measured for as many systems by
the NOFS program, including seven systems for which both CTIOPI and
NOFS measured parallaxes.  The CTIOPI results represent all completed
parallaxes for WDs on the program, including those beyond the 25 pc
horizon of interest.  A subset of the CTIOPI targets are members of
the 15 pc Astrometric Search for Planets Encircling Nearby Stars
\citep[ASPENS,][]{2003AAS...203.4207K} initiative.  These targets
typically have $\sim$10$+$ years of data and will be continually
monitored as long as the program will allow.  The NOFS parallaxes only
include those within the 25 pc horizon of interest.  A much larger
sample of WD parallaxes from NOFS at all distances will be included in
an forthcoming publication.

The combined astrometric efforts add 50 WD systems to the 25 pc sample
(27 from CTIOPI, 20 from NOFS, and 3 measured by both programs).
Also, we spectroscopically confirmed a previously unknown WD companion
to a main-sequence dwarf with a {\it Hipparcos} parallax placing the
system within 25 pc.  Finally, we confirm proximity for two WDs whose
trigonometric parallaxes were also recently determined by the {\it
  Tycho Gaia} astrometric solution
\citep[TGAS,][]{2016A&A...595A...4L}.  Thus, a total of 53 new systems
are added to the 25 pc WD sample that previously consisted of 126
systems reliably within that volume --- a 42\% increase.  The complete
sample of WDs within 25 pc can be found at
\url{http://www.DenseProject.com}.

\section{Observations and Data}
\label{sec:obsdat}

\subsection{Photometry}
\label{subsec:phot}

\subsubsection{Optical $BVRI$ Photometry}
\label{subsubsec:vriphot}

Standardized photometric observations were carried out at three
separate telescopes.  The SMARTS 0.9m telescope at CTIO was used
during CTIOPI observing runs when conditions were photometric.  A
Tektronics 2K $\times$ 2K detector was used in region-of-interest mode
centered on the central quarter of the full CCD producing a field of
view (FOV) of 6$\farcm$8 $\times$ 6$\farcm$8.  The SMARTS 1.0m
telescope at CTIO was used with the Y4KCam 4K $\times$ 4K imager,
producing a 19$\farcm$7 $\times$ 19$\farcm$7 FOV.  Finally, the
Ritchey 40-in telescope at USNO Flagstaff Station was used with a
Tektronics 2K $\times$ 2K detector with a 20$\farcm$0 $\times$
20$\farcm$0 FOV.  Calibration frames (biases, dome and/or sky flats)
were taken nightly and used to perform basic calibrations of the
science data using standard IRAF packages.  Standard stars from
\citet{1982PASP...94..244G} and \citet{1992AJ....104..340L,
  2007AJ....133.2502L, 2013AJ....146..131L} were taken nightly through
a range of airmasses to calibrate fluxes to the Johnson-Kron-Cousins
system and to calculate extinction corrections.  In general, aperture
photometry was performed on both standard stars and target stars using
a 14\arcsec~diameter.  For crowded fields, faint targets, and recent
observations, once a PSF pipeline was in place, PSF photometry was
conducted using either the {\it DAOPhot} \citep{1987PASP...99..191S}
or the {\it PSFEx} \citep{2011ASPC..442..435B} algorithm.  A subset of
data were compared using both PSF algorithms and no significant
systematic offset was seen.  While three separate Johnson-Kron-Cousins
$VRI$ filter sets were used between the three telescopes, comparisons
were made of dozens of CTIOPI targets mutually observed with all
filter sets.  Any systematic variation inherent in the filter set
differences once standardized is well below our nominal magnitude
error of 0.03 mag.

For the CTIOPI program, relative brightnesses were also recorded for
the parallax target (hereafter referred to as the ``PI'' star)
compared to the astrometric reference field stars in the filter used
for the astrometry as part of the CTIOPI reduction pipeline.  From
these data, we gauge whether the PI star shows any variability.  If
any of the reference stars show variability above $\sim$2\%, they are
removed from the variability analysis.  This analysis was not
performed for the NOFS targets as it was not part of the reduction
pipeline.  Photometry values are given in Table \ref{tab:pht}, where
columns (1) and (2) give WD and alternate names, respectively.
Columns (3-10) give the Johnson $BV$, Kron-Cousins $RI$, and
corresponding number of observations in each filter.  Columns (11-14)
give the filter of parallax observations (hereafter referred to as the
parallax filter) and PI star photometric standard deviation in that
filter as a gauge for variability as well as the number of nights and
frames used for the variability analysis.  Columns (15-17) give the
$JHK_s$ photometry values and corresponding errors on the 2MASS
photometric system.  Finally, column (18) contains any notes.

\subsubsection{NEWFIRM $JHK$ Photometry}
\label{subsubsec:newfirm}

Near-IR $JHK_s$ photometry was collected for WD 0851$-$246, at the CTIO
4.0-m Blanco telescope using the NOAO Extremely Wide-Field Infrared
Imager \citep[NEWFIRM,][]{2004SPIE.5492.1716P} during an engineering
night on 2011.27 UT.  NEWFIRM is a 4K $\times$ 4K InSb mosaic that
provides a 28\arcmin~$\times$ 28\arcmin~FOV on the Blanco telescope.
Raw data were processed using the NEWFIRM science reduction pipeline
and retrieved from the NOAO science archive as fully processed,
stacked images.

Relative photometry was performed using the 2MASS catalog to
standardize the images.  Frames were checked to identify where
saturation occurs and comparison stars were selected to have high
signal-to-noise yet below saturation.  A total of 68 comparison stars
were used for each frame with 2MASS magnitudes ranging from
12.61-14.42, 12.24-13.99, and 12.14-13.90 for $J$, $H$, and $K_s$,
respectively.  The NEWFIRM filters are on the MKO system so the
comparison stars were transformed to the MKO system using the
methodology of \citet{2001AJ....121.2851C}\footnote{The MKO
  transformations were not included in \citet{2001AJ....121.2851C} but
  were added later and available at
  \url{http://www.astro.caltech.edu/~jmc/2mass/v3/transformations/}.}.
Instrumental PSF photometry was extracted using {\it PSFEx} for the
comparison stars and the target.  A least-squares fit was used to
determine the offset between instrumental $J$ and MKO $J$.  A similar
approach was used to determine the MKO $J-H$ and $J-K$ colors.
Photometry values and errors are listed in Table \ref{tab:pht} and are
italicized to distinguish them from other $JHK_s$ values on the 2MASS
photometric system.

\subsubsection{Catalog Photometry}
\label{subsubsec:catphot}

Additional photometry values were extracted from the Sloan Digital Sky
Survey (SDSS) DR12 \citep{2015ApJS..219...12A}, 2MASS, and the UKIRT
Infrared Sky Survey (UKIDSS) DR9 Large Area Survey, when available.
The UKIDSS project is outlined in \citet{2007MNRAS.379.1599L}. UKIDSS
uses the UKIRT Wide Field Camera
\citep[WFCAM;][]{2007A&A...467..777C}. The photometric system is
described in \citet{2006MNRAS.367..454H}, and the calibration is
described in \citet{2009MNRAS.394..675H}. The science archive is
described in \citet{2008MNRAS.384..637H}.  UKIDSS magnitudes were
transformed to the 2MASS system using the transformations of
\citet{2009MNRAS.394..675H}.  These transformed values are listed in
Table \ref{tab:pht}.  We do not tabulate the photometry extracted from
SDSS DR12 as those are readily available via the SDSS archive.

\subsection{Spectroscopy}
\label{subsec:spec}

\begin{figure}[t!]
\centering
\includegraphics[angle=0, trim={1cm 0cm 0cm 1.5cm},
  width=0.50\textwidth]{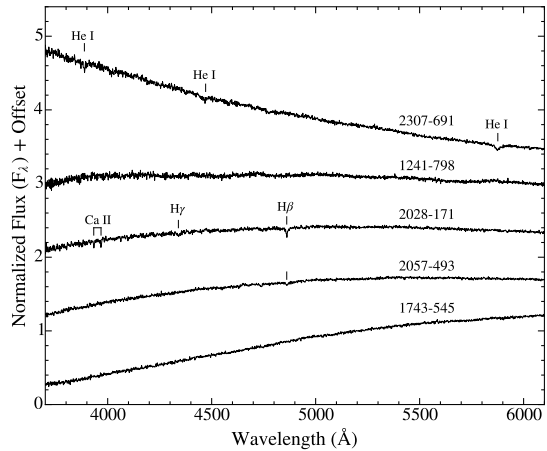}
\caption{SOAR $+$ Goodman confirmation spectra for newly discovered
  nearby WDs, WD 2307$-$691 (DB), WD 1241$-$798 (DC), WD 2028$-$171
  (DAZ), WD 2057$-$493 (DA) and WD 1743$-$545 (DC).  Spectra are
  normalized at 5200 \AA~and offset for clarity.  Supplemental data
  file of these spectra is available in the online journal.}
\label{fig:newspec}
\end{figure}

Two WDs presented here (WD 1743$-$545 and WD 2057$-$493) are newly
discovered nearby WDs identified during a spectroscopic survey of WD
candidates in the southern hemisphere (Subasavage et al. in
preparation) taken from the SUPERBLINK catalog
\citep{2005AAS...20715001L}.  A third WD included here (WD 2307$-$691)
was previously unclassified, yet is a common proper motion companion
to a {\it Hipparcos} star within 25 pc (HIP 114416).  A fourth WD (WD
2028$-$171) was suspected to be a WD by the authors based on a trawl
of the New Luyten Two Tenths (NLTT) catalog
\citep{1979nlcs.book.....L}.  Finally, A fifth WD (WD 1241$-$798) was
first spectroscopically identified as a WD by
\cite{2008AJ....136..899S} but with an ambiguous spectral type of
DC/DQ.  The SOAR 4-m telescope with the Goodman spectrograph was used
for spectroscopic follow-up as part of a larger spectroscopic campaign
to identify nearby WDs to be released in a future publication.
Observations were taken with a 600 lines-per-mm VPH grating with a
1$\farcs$0 slit width to provide 2.1 \AA~resolution in wavelength
range of 3600\AA$-$6200\AA.  The slit was rotated to the parallactic
angle to prevent any color-differential loss of light.  For WD
1241$-$798, the spectrum was taken during an engineering night and
only quartz lamp flats were taken.  The undulations seen in the
spectrum correlate with structure of the quartz lamp and are thus not
real.  For these spectra and throughout this work, we adopt the WD
spectral classification system of \citet{1983ApJ...269..253S}.  In
brief, DA WDs contain Balmer features, DB WDs contain helium features,
DC WDs are featureless, and DZ and DQ show calcium and carbon
features, respectively, and are devoid of hydrogen and helium
features.  If metal features are present as well as hydrogen and/or
helium features, the dominant species is listed first, i.e., DAZ is a
hydrogen-dominant atmosphere with traces of calcium.  The spectra for
WD 1241$-$798 (DC), WD 1743$-$545 (DC), WD 2028$-$171 (DAZ), WD
2057$-$493 (DA), and WD 2307$-$691 (DB), normalized at 5200\AA, are
shown in Figure \ref{fig:newspec}.

\subsection{Astrometry}
\label{subsec:astro}
\subsubsection{CTIOPI Astrometry}
\label{subsubsec:ctioast}

Trigonometric parallax data acquisition and reduction techniques for
the CTIOPI program are discussed fully in \citet{2005AJ....129.1954J}.
In brief, the instrument setup and basic data calibrations are
identical to those used for photometric observations (i.e., the SMARTS
0.9m telescope coupled with the central quarter of a Tektronics 2K
$\times$ 2K detector).  A parallax target's reference field is
determined upon first observation.  We use one of the
Johnson-Kron-Cousins $VRI$ filters, selected to optimize the signal on
the PI star and reference stars (the parallax filter), as well as to
keep exposure times greater than $\sim$30 seconds and less than
$\sim$600 seconds, when possible.  Rapid variations in the atmosphere
are not ``smoothed'' out in short exposures and thus degrade the
centroids.  Integrations longer than 600 seconds are taxing to the
program as a whole as we typically have $\sim$300-500 active targets
at any given time.  In general, five frames are taken at each epoch
for shorter-exposure targets and three frames per epoch for
600-second-exposure targets.  WD 0222$-$291 (see Section
\ref{subsec:comments}) is one of the faintest targets in the parallax
filter observed during the program and required $\sim$900 second
integrations.  In this case, only two frames per epoch (at most) were
taken.

When possible, the PI star is placed as close to the center of the CCD
as possible, to allow a roughly circular distribution of reference
stars around the PI star.  Preference in reference star selection is
given to better exposed (though not near saturation), isolated stars
that are near to the PI star on the CCD.  A minimum of five to ten
reference stars is preferred, up to a maximum of $\sim$20 reference
stars that are sufficiently exposed.  Diminishing returns are quickly
realized if more reference stars are used, often because they are
poorly exposed.  Exceptions are made for sparse fields when necessary,
typically at the expense of a slightly degraded parallax solution.
Once determined, the pointing to each field is repeated to better than
$\sim$5-10 pixels throughout the duration of the parallax observations
to minimize any CCD distortion effects on the final astrometric
solution.  Generally, observations are limited to within an hour angle
of 2 hours, with most observations taken within $\sim$30 minutes of
the meridian.

Astrometric reductions for each parallax target are performed/updated
routinely to monitor progress and identify any problematic datasets.
Centroids for the reference field and PI stars are determined for all
frames in the parallax filter using {\it SExtractor}
\citep{1996A&AS..117..393B} outputs XWIN\_IMAGE and YWIN\_IMAGE.
Corrections for differential color refraction (DCR) are performed
using the $VRI$ photometry of all the reference stars and PI star.  A
least-squares reduction via the Gaussfit program
\citep{1988CeMec..41...39J} is performed, assuming the reference star
grid has $\Sigma\pi_i =$ 0 and $\Sigma\mu_i =$ 0, where $\pi$
and $\mu$ are parallax and proper motion, respectively.  Once a
relative parallax is obtained for the PI star, a correction to
absolute parallax is determined by estimating photometric distances
via the relations of \citet{2004AJ....128.2460H} to the reference
stars (assuming all are dwarfs).

For three stars, WD 1241$-$798, WD 1620$-$391, and WD 1917$-$077, that
reside within 20$^\circ$ of the Galactic Plane and exhibit
significantly reddened fields, we performed a more rigorous
calculation to correct from relative to absolute parallax.  The
methodology is discussed fully in \citet{2016AJ....152..118H} but in
brief, the reddening of the field is estimated using E($B-V$) taken
from \citet{2011ApJ...737..103S}.  Near infrared 2MASS
photometry is used to determine which reference stars are likely
giants, and the $V-I$ color is then adjusted iteratively as a distance
estimate converges for each reference star, assuming a dwarf or giant.
In the case of WD 1241$-$798, the correction to absolute parallax is
determined to be 1.1 $\pm$ 0.3 mas and for the latter two, the
correction is determined to be 1.6 $\pm$ 0.2 mas.

\subsubsubsection{Cracked $V$ Filter}
\label{subsubsubsec:cracked}

A complete discussion of the cracked $V$ filter issue can be found in
\citet{2009AJ....137.4547S}; here we give a synopsis.  Because of a
damaged Tek 2 $V$ filter (referred to as $oV$) that occurred in early
2005, the CTIOPI program used a comparable $V$ filter (referred to as
$nV$) from 2005 to mid-2009.  The astrometry is affected by this
change because the passbands were slightly different.  It was
determined empirically that trigonometric parallax determinations are
sound if at least $\sim$1-2 years of data are available both before
and after the filter switch.  However, subtle signals from a
perturbing companion, would not be reliable.  In 2009, it was
determined that the crack, near the corner of the filter did not
impact the FOV of the CTIOPI data, as only the central quarter of the
CCD is used.  Thus, a switch back to the original $V$ ($oV$) was
completed in mid-2009.

For the targets, particularly in the ASPENS 15 pc sample where we want
to probe for subtle astrometric signatures in the residuals, the $nV$
data are omitted from the reductions presented here.  Otherwise,
reductions that include both $V$ filter data are noted in Table
\ref{tab:ctioast}.  In the case of WD 1241$-$798, no new data were
taken after 2009 and only a year of data were taken with $oV$ prior to
2005.  Thus, only the $nV$ data are used to determine the astrometric
results presented here.

\subsubsection{NOFS Astrometry}
\label{subsubsec:nofsast}

A thorough discussion of the NOFS astrometric reductions can be found
in \citet{1992AJ....103..638M} and \citet{2002AJ....124.1170D} with
procedural updates described in Dahn et al.~(2017, in preparation).
Briefly, astrometric data have been collected with the Kaj Strand
61-in Astrometric Reflector \citep{1964S&T....27..204S} using three
separate CCDs over the multiple decades that NOFS has measured stellar
parallaxes.  Initially, a Texas Instruments (TI) 800 $\times$ 800
(TI800) CCD, followed by a Tektronics 2048 $\times$ 2048 (Tek2K) CCD,
and most recently an EEV (English Electric Valve, now e2v) 2048
$\times$ 4096 (EEV24) CCD were used.  The latter two cameras are still
in operation at NOFS for astrometric work and were used for all but
two of the NOFS parallaxes presented here.  The TI800 CCD was used to
measure the parallaxes for WD 0213$+$396 and WD 1313$-$198.  A total
of four filters were used for astrometric work. ST-R (also known as
STWIDER) is described in detail by \citet{1992AJ....103..638M}, and is
centered near 700 nm with a FWHM of 250 nm.  A2-1 is an optically flat
interference filter centered near 698 nm with a FWHM of 172 nm.  I-2
is an optically flat interference filter centered near 810 nm with a
FWHM of 191 nm.  Z-2 is an optically flat 3 mm thick piece of Schott
RG830 glass that produces a relatively sharp blue-edge cutoff near 830
nm and for which the red edge is defined by the CCD sensitivity.  More
details on the filters can be found in Dahn et al.~(2017, in
preparation).

Reference stars are selected during initial setup, typically with more
selections than required.  Centroids for the reference field and PI
star occur on-the-fly as data are collected using the centroiding
algorithm of \citet{1983AJ.....88.1489M}.  A comparison of this
algorithm and that of {\it SExtractor} as used for CTIOPI, using
several parallax fields, show them to produce comparable results.
Corrections for DCR were determined based on the $V-I$ colors and
applied to the PI and reference star centroids prior to the
astrometric solution.  An astrometric solution is then calculated to
give relative parallax and proper motion.

The correction to absolute parallax is determined using the
methodology of \citet{2016AJ....152..118H} and the same as that
described for the reddened cases in the CTIOPI program.  Corrections
for most of the targets presented here do not require the use of 2MASS
photometry to determine reference stars likely giants vs.~dwarfs, as
reddening is minimial.  The correction to absolute for WD 1821$-$131
was not determined in this manner because the
\citet{2011ApJ...737..103S} determination of E($B-V$) $=$ 13.9 for
this field and thus, giant/dwarf differentiation was very unreliable.
Instead, we adopt a nominal correction with an inflated error of 1.0
$\pm$ 0.3 mas for this target.

\section{Astrometry Results}
\label{sec:astres}

CTIOPI astrometric results for WD systems (and companions when
available) are presented in Table \ref{tab:ctioast}.  Columns (4)-(9)
list the filter used for parallax observations, the number of seasons
the PI star was observed, the total number of frames used in the
parallax reduction, the time coverage of the parallax data, and the
number of reference stars used.  The `c' in column (5) signifies that
the observations were continuous throughout every season within the
time coverage.  The `s' signifies that observations were scattered
such that there is at least one season with only one night's data (or
no data for an entire season).  In some cases, mostly because of the
cracked $V$ filter problem discussed in Section
\ref{subsubsubsec:cracked}, the `g' signifies a significant gap
(multiple years) in the observations.  Columns (10)-(12) list the
relative parallax, correction to absolute, and the absolute parallax.
The proper motions and position angles quoted in columns (13) and (14)
are those measured with respect to the reference field (i.e.,
relative, not corrected for reflex motion due to the Sun's movement in
the Galaxy).  The tangential velocities quoted in column (15) are not
corrected for solar motion.  For the ASPENS targets that were
published by \citet{2009AJ....137.4547S}, continual monitoring over
the past $\sim$6 years has provided significant additional data, and
thus the astrometric results presented here supersede those
previously published.  The mean error on the parallax for the CTIOPI
sample is 1.14 mas.

\begin{figure}[b!]
\centering
\includegraphics[angle=0, trim={1cm 1cm 0cm 2.2cm},
  width=0.50\textwidth]{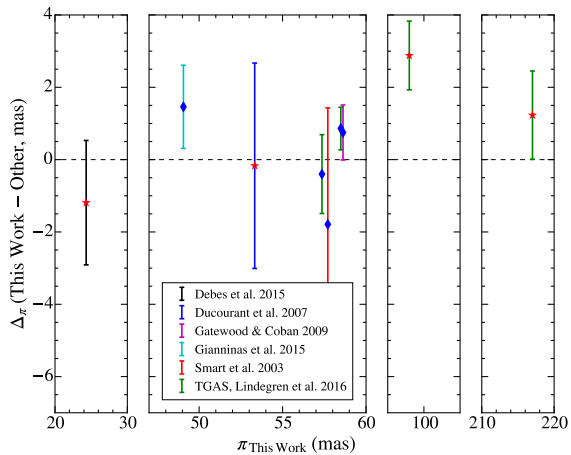}
\caption{Comparison of trigonometric parallax results with other
  authors' recent results, shown as $\Delta\pi$ (this work $-$
  other).  Comparisons with CTIOPI are shown as red stars and with
  NOFS as blue diamonds.}
\label{fig:compare}
\end{figure}

NOFS astrometric results for 25 pc WD systems are presented in Table
\ref{tab:nofsast}.  Columns (4)-(9) list the filter used for parallax
observations, the number of nights the PI star was observed, the total
number of frames in the astrometric reduction, the number of reference
stars used, and the time coverage and length of the parallax data.
Columns (10)-(14) list the relative parallax, correction to absolute,
the absolute parallax, and relative proper motion and position angle
(i.e., not corrected for solar motion).  Also in this case, the
tangential velocities quoted in column (15) are not corrected for
solar motion.  Finally, column (16) denotes which camera was used for
parallax observations.  The mean error on the parallax for the NOFS
sample is 0.49 mas, or roughly a factor of two better than that for
CTIOPI.  The enhanced accuracy is attributed to the astrometric
optimization of the NOFS 61-in Strand Reflector's optical design.

\begin{figure*}[!ht]
\centering
\includegraphics[angle=0, trim={0 0.5cm 2cm 3cm},
  width=0.95\textwidth]{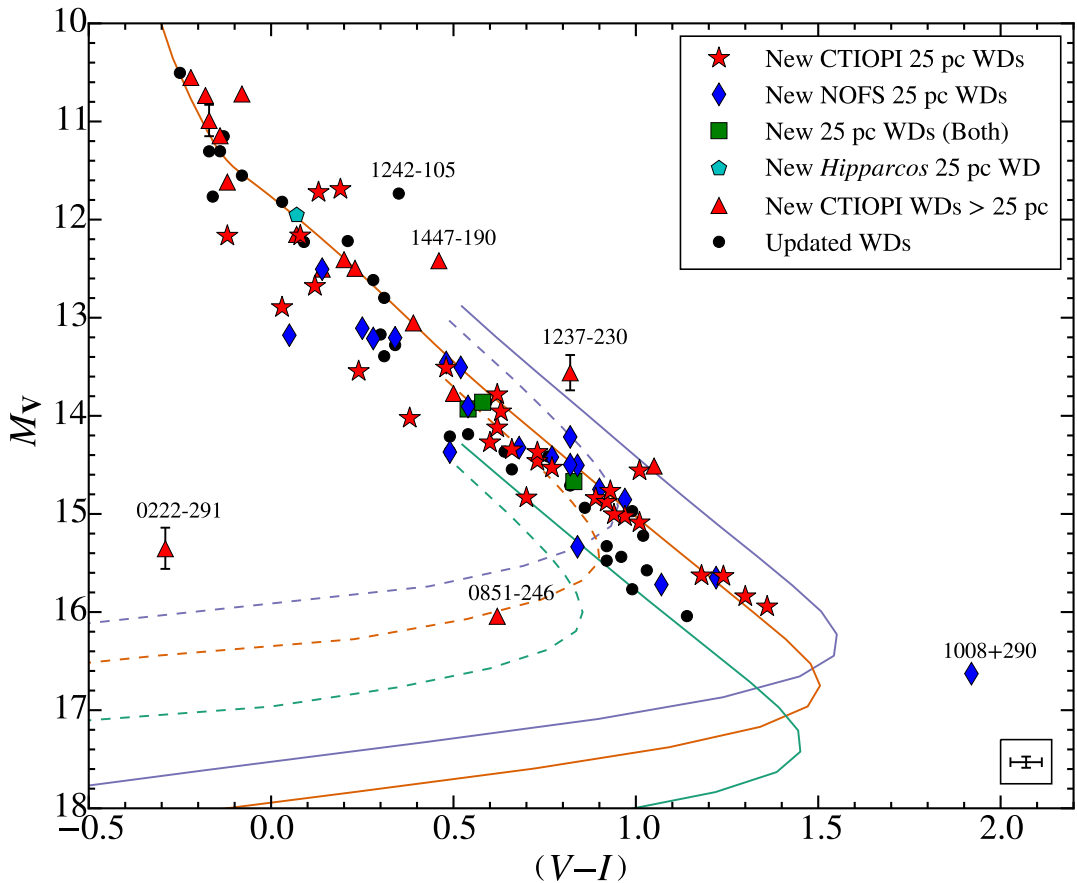}
\caption{Hertzsprung-Russell diagram for the objects with parallaxes
  presented here, separated by sample.  For the CTIOPI sample, objects
  beyond 25 pc are included as red triangles.  Objects labeled by WD
  name are discussed in Section \ref{subsec:comments}.  A
  representative error bar is encased in the lower right corner of the
  plot.  Error bars are plotted explicitly for the three objects whose
  $M_V$ errors are larger than 0.15 mag.  Curves represent atmospheric
  model tracks for three values of \logg, 7.5 ({\it purple}), 8.0
  ({\it orange}), and 8.5 ({\it green}).  Solid curves are pure-H
  models and the dashed curves are mixed atmosphere He$+$H models of
  \citet{1995ApJ...443..764B}, with improvements discussed in
  \citet{2009ApJ...696.1755T}, with log [He/H] = 2.0.  All other
  reasonable mixed atmosphere models fall between these two curves of
  a given \logg.  For all models other than the pure-H model at \logg
  = 8.0, curves are plotted for $T_{\rm eff}$ = 7,000K and cooler.}
\label{fig:hrdiag}
\end{figure*}

In Figure \ref{fig:compare} we compare astrometric results with
recently published works for the few overlapping targets.  The error
bars represent both programs' formal parallax errors added in
quadrature for a given target.  There are no obvious systematic
differences with either CTIOPI or NOFS samples.

Figure \ref{fig:hrdiag} shows a H$-$R diagram for the astrometric
samples presented here.  Objects labeled by WD name are discussed in
detail in Section \ref{subsec:comments}.  Briefly, we find three WDs
(WD 1242$-$105, WD 1447$-$190, and WD 1237$-$230) that are
overluminous and good candidates for being unresolved multiple systems
solely based on luminosity.  WD 1008$+$290 is a peculiar He-rich DQ WD
with exceptional Swan band absorption
\citep[e.g.,][]{2012ApJS..199...29G} such that the measured $V$
magnitude, whose bandpass encompasses a portion of this absorption, is
affected and appears fainter than if the absorption was not present.
Finally, both LHS 2068 (WD 0851$-$246) and LHS 1402 (WD 0222$-$291)
are very cool WDs that appear to display collision-induced absorption
(CIA) by H$_2$ molecules \citep{1994ApJ...424..333S,
  1998Natur.394..860H, 1999ApJ...511L.107S}.  CIA opacity is induced
by collisions, and thus requires high atmospheric pressures. High
atmospheric pressures are reached at much higher effective
temperatures in He-rich atmospheres that also contain molecular
hydrogen, because of the relative transparency of He, than in pure-H
atmospheres.  Therefore, CIA manifests itself at higher effective
temperatures, and thus higher luminosities.  This effect is shown in
Figure \ref{fig:hrdiag} compared to the effect of CIA in pure-H
atmospheres.  In the case of LHS 2068, a mixed atmosphere He$+$H is
required to allow for CIA to be present at its luminosity (discussed
further in Section \ref{subsec:comments}).  In the case of LHS 1402,
none of the current models compare adequately to the observed data
(also discussed further in Section \ref{subsec:comments}).  As
expected, the majority of new 25 pc WD members are cooler and
intrinsically dimmer than their hotter counterparts and often missed
in magnitude-limited surveys.


\section{Analysis}

\subsection{Modeling of Physical Parameters}
\label{subsec:model}

\begin{figure*}[t!]
\centering
\includegraphics[angle=0, trim={3cm 2cm 3cm 2cm},
  height=0.65\textheight]{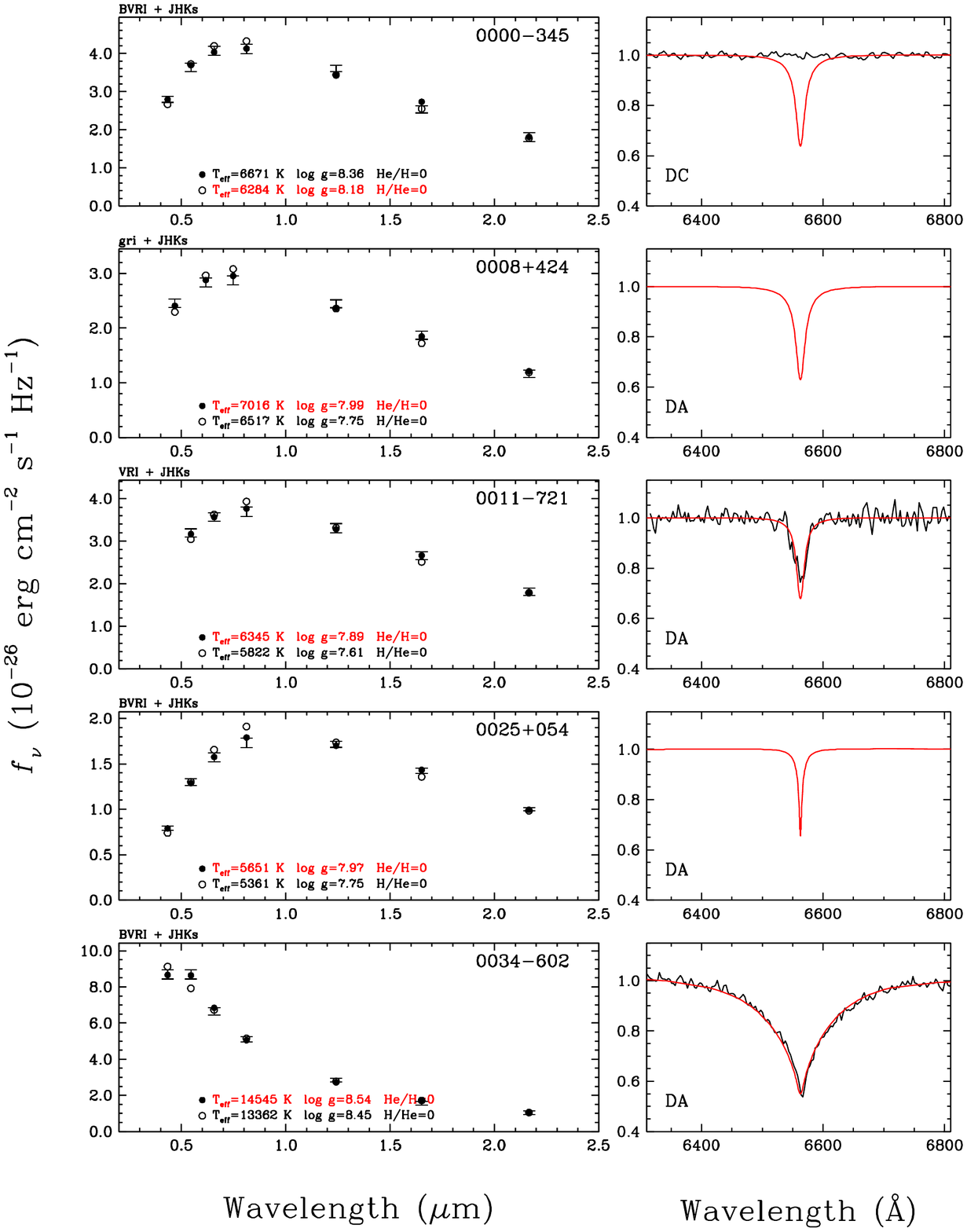}
\caption{Representative plots of the model fits to the SEDs where
  error bars represent measured values and circles represent model
  values (filled for pure-H and open for pure-He).  Adopted model
  parameters are shown in red text in the left panels.  Plots of all
  model fits for the sample are available in the online material.
  Among the online plots, note that both WD 0038$-$226 and WD
  0851$-$246 are fit using mixed-atmosphere He$+$H models.  All
  photometric measures that were excluded from the fit are shown as
  red error bars and described in Section \ref{subsec:comments}.
  Model spectra in the right panels are not fits, but merely derived
  from the adopted model atmospheres assuming pure-H.  Comparisons
  with measured spectra (black lines), when available, are shown and
  serve as consistency checks for the DA WDs and to highlight any
  anomalies.  Fits for DQ and DZ are plotted separately and, in these
  cases, spectral fitting is performed as described in Section
  \ref{subsec:model}.}  \notetoeditor{Please show the first page of
  the allfits_photo.pdf and include the full multi-page version as
  online material.  Please also append both 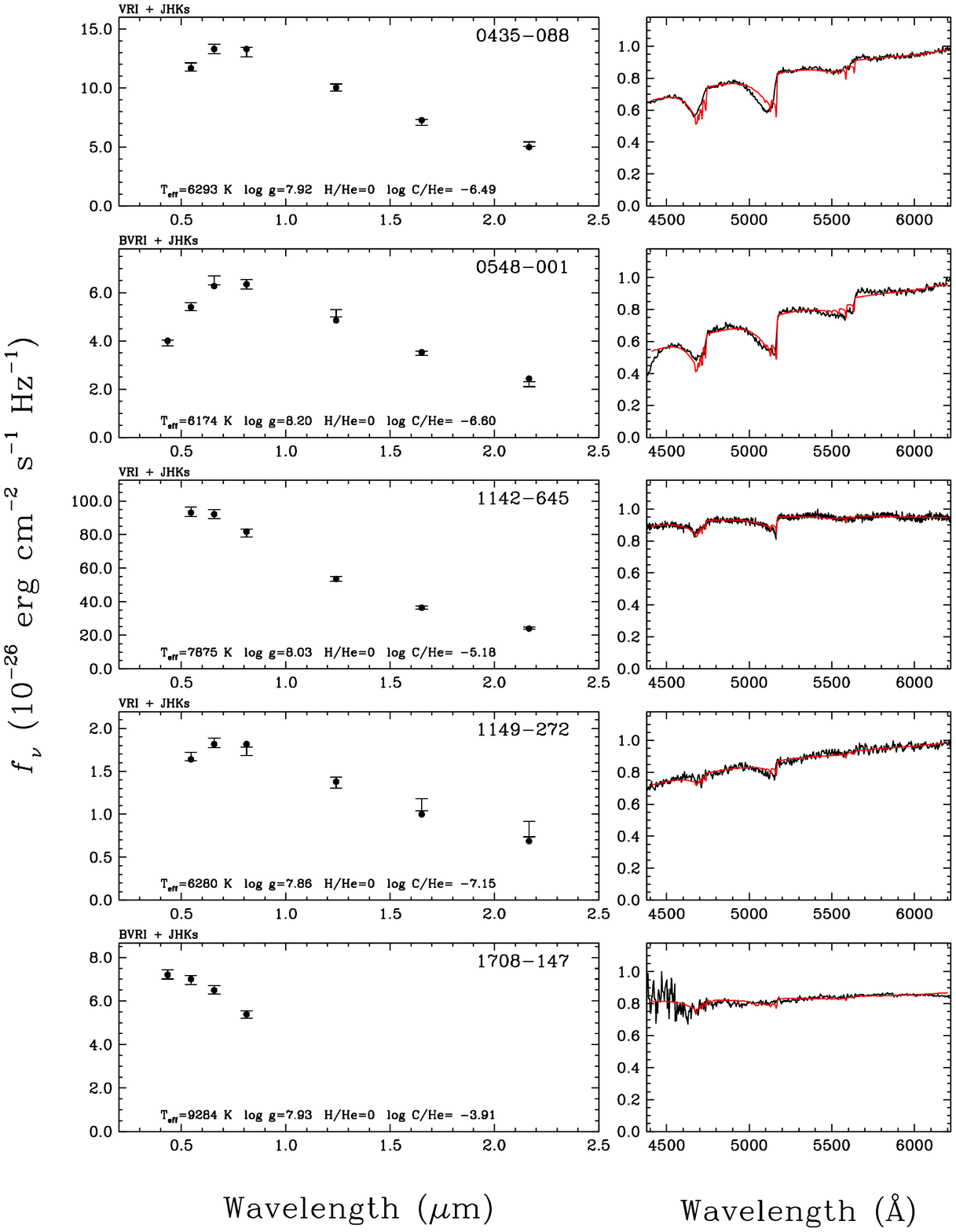 and
  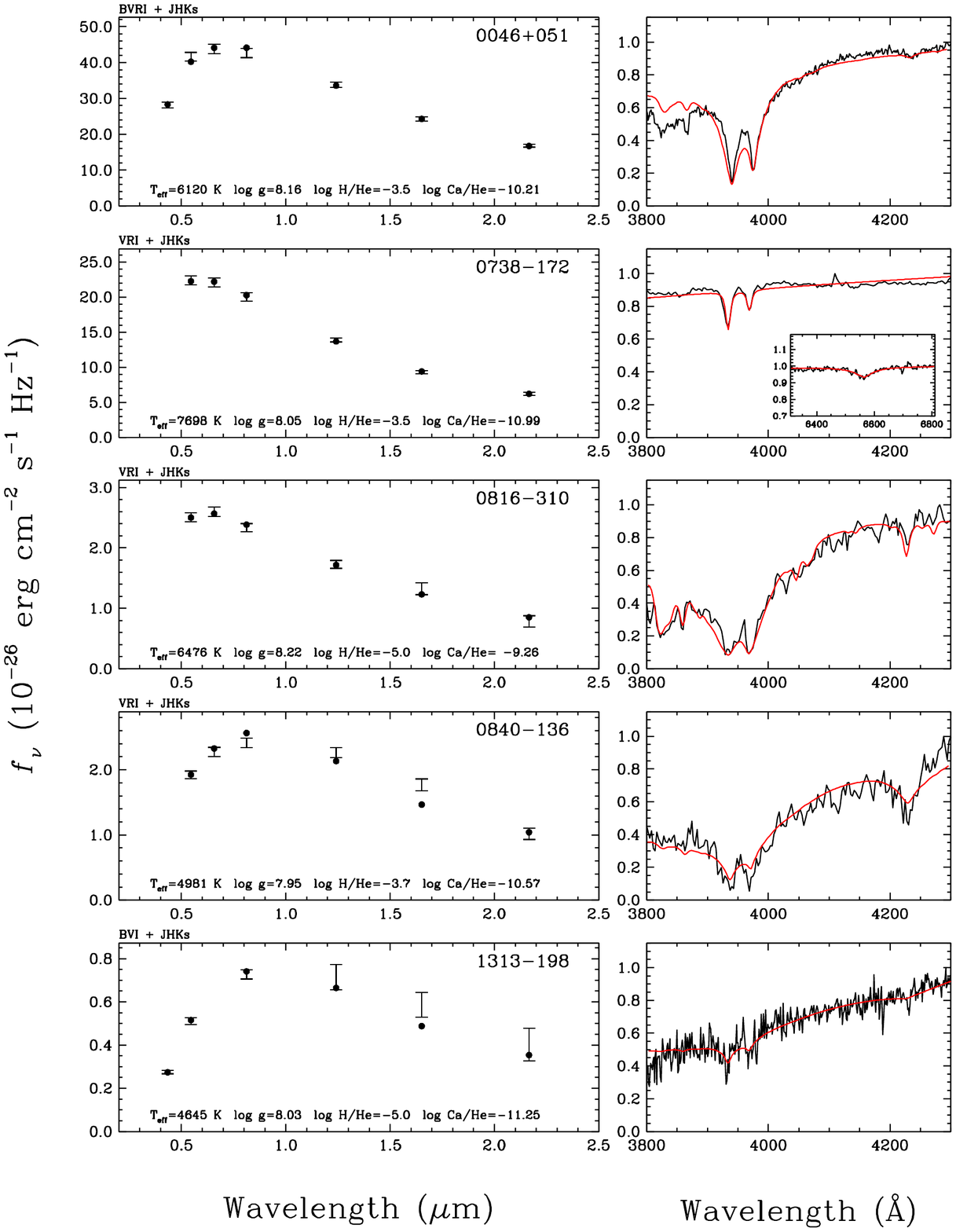 as part of this figure in the online material.}
\label{fig:fitplots}
\end{figure*}

To better understand the physical nature of WDs, which expands to
provide clues into topics such as stellar evolution and progenitor
populations, for example, we perform atmospheric modeling analyses of
all of our targets.  Atmospheric modeling procedures of the WDs are
identical to those presented in \citet{2009AJ....137.4547S}.  Briefly,
optical/near-IR magnitudes are converted into fluxes using the
calibration of \citet{holberg06} and compared to the spectral energy
distributions (SEDs) predicted by the model atmosphere calculations
originally described in \citet[][and references
  therein]{1995ApJ...443..764B} with improvements discussed in
\citet{2009ApJ...696.1755T}.  The observed flux, $f_{\lambda}^m$, is
related to the model flux by the equation
\begin{equation}
f_{\lambda}^m= 4\pi~(R/D)^2~H_{\lambda}^m
\end{equation}
\noindent
where $R/D$ is the ratio of the radius of the star to its distance
from Earth, $H_{\lambda}^m$ is the Eddington flux (dependent on
$T_{\rm eff}$, \logg, and atmospheric composition) properly averaged
over the corresponding filter bandpass, and $\pi$ is the mathematical
constant (elsewhere throughout this paper, $\pi$ refers to the
trigonometric parallax angle).  Our fitting technique relies on the
nonlinear least-squares method of Levenberg-Marquardt
\citep{pressetal92}, which is based on a steepest descent method. The
value of $\chi ^2$ is taken as the sum over all bandpasses of the
difference between both sides of Equation (1), weighted by the
corresponding photometric uncertainties.  Only $T_{\rm eff}$ and [$\pi
  (R/D)^2$] are free parameters (though we allow \logg to vary as
described below) and the uncertainties of both parameters are obtained
directly from the covariance matrix of the fit.  The main atmospheric
constituent (hydrogen or helium) is determined by the presence of
H$\alpha$ from spectra published in the literature (references listed
in Table \ref{tab:parameters}) or by comparing fits obtained with both
compositions.

We start with \logg = 8.0 and determine $T_{\rm eff}$ and [$\pi
  (R/D)^2$], which combined with the distance $D$ obtained from the
weighted mean trigonometric parallax measurement yields directly the
radius of the star $R$. The radius is then converted into mass using
evolutionary models similar to those described in \citet{fon01}, but
with C/O cores, $q({\rm He})\equiv \log M_{\rm He}/M_{\star}=10^{-2}$
and $q({\rm H})=10^{-4}$ (representative of hydrogen-atmosphere WDs),
and $q({\rm He})=10^{-2}$ and $q({\rm H})=10^{-10}$ (representative of
helium-atmosphere WDs).\footnote{see
  \url{http://www.astro.umontreal.ca/~bergeron/CoolingModels/}.}  In
general, the \logg value obtained from the inferred mass and radius
($g=GM/R^2$) will be different from our initial guess of \logg = 8.0,
and the fitting procedure is thus repeated until an internal
consistency in \logg is reached. The parameter uncertainties are
obtained by propagating the error of the trigonometric parallax
measurements into the fitting procedure.

Physical parameter determinations for the DQ and DZ WDs are identical
to the procedures outlined in \citet{2005ApJ...627..404D,
  2007ApJ...663.1291D}.  Briefly, the photometric SED provides a first
estimate of the atmospheric parameters with an assumed value of metal
abundances using solar abundance ratios.  The optical spectrum is fit
to better constrain the metal abundances and to improve the
atmospheric parameters from the photometric SED.  This procedure is
iterated until a self-consistent photometric and spectroscopic
solution is reached.

Results of the atmospheric modeling are tabulated in Table
\ref{tab:parameters}.  Given the nominal uncertainties, we round the
values for effective temperature and corresponding error to 10 K.
Representative plots of the model fits of the SEDs are shown in Figure
\ref{fig:fitplots}, with the complete set of plots for the entire
sample made available in the online material.

\subsection{Comments on Individual Systems}
\label{subsec:comments}

{\bf WD 0000$-$345} belongs to an unusual class of objects whose SEDs
are better fit with pure-H atmospheric models, yet show no Balmer
spectral features.  This class was first identified by
\citet{1997ApJS..108..339B} and more recently discussed in
\citet{2012ApJS..199...29G}, who suggest that strong magnetic fields
could be the cause of the discrepancy.  In fact, WD 0000$-$345 was
initially classified as a magnetic WD by \citet{1996A&A...311..572R}
and later classified as a DC by \citet{1997ApJS..108..339B}.  Circular
polarization studies by \citet{2001MNRAS.328..203S} show no detectable
magnetic field, and thus we classify this object as a DC.  We adopt
the pure-He atmospheric models to remain consistent with model choice
based on spectroscopy, where available.

{\bf WD 0127$-$311} was initially classified as a magnetic WD by
\citet{1996A&A...311..572R} yet circular polarization studies by
\citet{2001MNRAS.328..203S} show no detectable magnetic field.  While
the spectrum shows what appears to be He \textsc{I} 5876 \AA,
indicative of a DB, \citet{2001MNRAS.328..203S} argue that other
helium features at 4472 \AA~and 6678 \AA~should be present for a wide
range of temperatures, yet are absent here.
\citet{2001MNRAS.328..203S} suggest the object could be a DZ with the
Na \textsc{I} D lines (5890 and 5896 \AA) but the spectral profile
seen is asymmetric, unlike what is expected for the Na lines.  We
favor the former interpretation and have adopted a spectral type of DB
for this object.  Given that it is a rather cool DB, the He \textsc{I}
5876 \AA~line will be the last feature seen as the WD cools to a DC
spectral type.  In fact, this DB is of a very similar effective
temperature (10,910 K) to the newly discovered DB, WD 2307$-$691 (also
10,910 K), whose spectrum is shown in Figure \ref{fig:newspec} and
also displays a prominent He \textsc{I} 5876 \AA~ feature and little
else.

{\bf WD 0222$-$291} was first discovered to be a high proper motion
star by \citet{1972PMMin..32....1L} \citep[LP 885$-$57; LHS
  1402,][]{1979lccs.book.....L} with $\mu =$ 0.501 \arcyear at
position angle 96.3$^\circ$.  It was later observed spectroscopically
by \cite{2001Sci...292..698O} and found to be a featureless ultracool
WD with strong infrared CIA that extends into the optical bandpasses,
thereby giving it significantly blue optical colors.
\cite{2001Sci...292..698O} postulated that WD 0222$-$291 was direct
evidence of galactic halo dark matter (based on kinematics) along with
the other 37 WDs presented in that publication.  The claim of these
objects being halo members was contested based on the relatively young
ages of these systems \citep{2003ApJ...586..201B}.  However, there
remained an ambiguity with WD 0222$-$291 as to whether the CIA was due
to collisions with hydrogen molecules only (pure-H atmosphere) or due
to collisions with neutral helium (mixed atmosphere He$+$H) that can
produce comparable CIA at higher temperatures.  The difference in
luminosity between the two scenarios is significant.  If a pure-H
atmosphere is assumed, WD 0222$-$291 would be one of the nearest WDs
to the Sun at 4.7 pc \citep{2003ApJ...586..201B}.  If the atmosphere
is dominated by helium with only trace amounts of hydrogen, then WD
0222$-$291 is considerably more distant at 20 - 25 pc
\citep{2004ApJ...601.1075S,2005ApJ...625..838B}.  The latter scenario
was preferred because the spectrum did not exhibit a broad absorption
feature near 8000 \AA~as expected in pure-H models (see Figure 5 of
\citealt{2005ApJ...625..838B}).

Here we confirm, with our trigonometric distance of 34.6 $\pm$ 3.5 pc,
that WD 0222$-$291 is indeed more consistent with a helium-dominant
atmosphere with the CIA being due to trace amounts of hydrogen in this
mixed atmosphere.  However, we do not arrive at a satisfactory fit
using any of the currently available models.  The object is
overluminous compared to predictions and, thus, could be a double
degenerate or the current models are inadequate (or both).  Therefore,
we refrain from listing any physical parameters for this WD.  In that
regard, WD 0222$-$291 can serve as an important empirical test case
for future atmospheric models.

We note that our parallax error is uncharacteristically large and we
attribute this to three causes.  First, this object is exceptionally
faint so that the signal-to-noise of our CTIO 0.9m astrometry images
is quite low.  Second, long integration times ($\sim$15 minutes per
frame, on occasion), produced images with less-than-circular contours
because of imperfect guiding.  Third, fewer frames were collected
because of the costly integration times and their impact on the rest
of the parallax program.  Nonetheless, the parallax result shows no
indication of systematics resulting from these effects and the error
is slightly better than 10\% of the parallax measured.  Indeed, over
the 6$+$ years that parallax data were collected, the parallax value
stabilized such that new data did not change the result, indicative of
a robust determination.

{\bf WD 0311$-$649} was classified as a DA by
\citet{2008AJ....136..899S} for which they estimated a distance of
21.0 pc.  The trigonometric parallax presented here implies a
significantly greater distance such that the implied mass, assuming a
single star, is remarkably low.  This object is likely an unresolved
double degenerate.

{\bf WD 0322$-$019} is shown by \citet{2003ApJ...596..477Z} to be a
double degenerate where both components show metal lines and are thus
DAZ spectral types.  In addition, \citet{2011MNRAS.413.2559F} identify
a magnetic field of $\sim$120 kG and model the system using a
preliminary parallax from USNO of 58.02 $\pm$ 0.44 mas.  The updated
parallax presented here, which includes several years of additional
data, is entirely consistent.

{\bf WD 0326$-$273} was confirmed by \citet{2003ApJ...596..477Z} to be
a double degenerate based on radial velocity variations of $H\beta$.
In addition, they find Ca II lines, but attribute the source as being
interstellar given the discrepant radial velocities compared to
$H\beta$.

{\bf WD 0552$-$041} is a weak DZ.  However, as discussed in
\citet{2012ApJS..199...29G}, the sharp Ca \textsc{II} H \& K lines
indicate a hydrogen-rich atmosphere.  If the atmosphere were
helium-rich, the atmospheric pressure would be greater, and thus the
Ca \textsc{II} line profiles would be much broader and shallower.
Therefore, we utilize a pure-H model to derive the physical parameters
in Table \ref{tab:parameters}.

{\bf WD 0802+387} has a recent trigonometric parallax determination by
\citet{2015MNRAS.449.3966G} of 47.6 $\pm$ 1.0 mas.  Thus, their
measure adds this target to the 25 pc sample, bringing the number of
reliable member systems to 126 prior to this work.  Our trigonometric
parallax presented here is entirely consistent and is included in our
number counts as an updated measure.  The weighted mean of these two
measures is used to calculate physical parameters shown in Table
\ref{tab:parameters}.


{\bf WD 0851$-$246} was first discovered to be a high proper motion
star by \citet{1974PMMin..37....1L} \citep[LP 844$-$26; LHS
  2068,][]{1979lccs.book.....L} with $\mu =$ 0$\farcs$630 yr$^{-1}$ at
position angle 78.0$^\circ$.  It has a common proper motion companion
(LHS 2067) whose spectral type is listed as sdM in
\cite{1995AJ....109..797K}.  The WD component was first
spectroscopically identified by \cite{2001ApJ...558..761R}, named CE
51 by those authors, during a follow-up spectroscopic campaign
targeting proper motion objects detected in the Cal{\'a}n-ESO survey
\citep{2001ApJS..133..119R}.  They determine that the WD is very cool
(2730 K), old (11.9 Gyr), and nearby (14.7 $\pm$ 0.3 pc).
\citet{2002AJ....124.1118S} conducted a radial velocity survey of wide
binaries with WD components, including this system, and found a
$V_{\rm rad} =$ 0.0 $\pm$ 15.8 km/sec for the sdM.  Based on the WD's
location in the H-R digram (Figure \ref{fig:hrdiag}), it is expected
that CIA is present and likely results from collisions with molecular
helium, thus, is a mixed atmosphere He$+$H.  The physical parameters
presented in Table \ref{tab:parameters} are derived from a mixed
atmosphere model with the fit shown in Figure \ref{fig:fitplots}.
This is the coolest WD presented here for which we were able to derive
reliable atmospheric parameters.  Notably, the mixed atmosphere fit
results in a significantly higher $T_{\rm eff}$ (3490 K) than that
found by \citet{2001ApJ...558..761R}.

Additionally, \cite{2009ApJ...696.2094K} observed this system with
Spitzer in the IRAC bandpasses, but stated that contamination from a
nearby bright star makes the photometry questionable.  As discussed in
Section \ref{subsubsec:newfirm}, we obtained $JHK_s$ photometry using
NEWFIRM on the CTIO Blanco 4-m.  The SED clearly shows excess in the
near-IR indicative of a cool, very red tertiary companion.  Follow-up
observations, in particular, near-IR spectroscopy, are required to
characterize the tertiary companion.  The mixed atmosphere fit shown
in Figure \ref{fig:fitplots} excludes the $JHK_s$ values; the optical
magnitudes are not affected by this very red unseen companion.

{\bf WD 1008$+$290} is a peculiar DQ WD with exceptional Swan band
absorption such that the measured $V$ magnitude, whose bandpass
encompasses a portion of this absorption, is affected and appears
fainter than if the absorption was not present.  Additionally, the
$V-I$ color is inflated for the same reason resulting in the WD's
displacement in Figure \ref{fig:hrdiag}.  We adopt a pure-He model
atmosphere simply because there was no satisfactory agreement with the
SED, regardless of model used; thus, the atmospheric parameters are
likely unreliable.

{\bf WD 1036$-$204} is another peculiar DQ WD with exceptional Swan
band absorption.  We adopt a pure-He model atmosphere simply because
there was no satisfactory agreement with the SED, regardless of model
used, thus, the atmospheric parameters are likely unreliable.

{\bf WD 1237$-$230} was first discovered to be a high proper motion
star by \citet{1972PMMin..30....1L} \citep[LP 853$-$15; LHS
  339,][]{1979lccs.book.....L} with $\mu =$ 1.102\arcyear at position
angle 219.9$^\circ$.  It was first classified as a DA WD by
\cite{1977ApJ...217L..59L}.  \cite{2007AJ....134..252S} obtained
optical $VRI$ photometry and combined it with $JHK_S$ from 2MASS to
model the SED and derive physical parameters.  One significant
discrepancy is the distance estimate derived (26.9 $\pm$ 4.5 pc)
compared to the trigonometric distance determined in this work (39.4
$\pm$ 3.6 pc).  The recent spectroscopic analysis of
\citet{2012MNRAS.425.1394K} showed this object to be a single-lined
binary, and they identify this object to be a candidate halo WD based
on the estimated distance.  Our trigonometric parallax further
strengthens the case for halo membership, as it has an extreme
tangential velocity of 202.4 \kmsec.  We see no evidence for
photocentric motion in the astrometric dataset, leaving open two
possibilities: (1) a system with two roughly equal-mass, equal
luminosity components, or (2) a system with a period short enough (P
$\lsim$ 1 year) to evade detection in our astrometric data.  Given the
overluminosity in the optical (see Figure \ref{fig:hrdiag}), such that
the components are of similar brightnesses, coupled with the
single-lined radial velocity variation, the system is likely composed
of a DA and a non-DA pair of WDs.

{\bf WD 1241$-$798} was classified as a DQ/DC by
\cite{2008AJ....136..899S} without further explanation.  The ambiguous
spectral classification arises because the authors obtained a red
spectrum from the Blanco 4-m at CTIO using the R-C Spectrograph.  With
a blue cutoff of $\sim$5500 \AA, no sharp features were present, yet
there were depressions very close to where the C$_2$ Swan bands are
expected but the detection is marginal.  To remove this discrepancy,
we acquired a confirmation spectrum using SOAR $+$ Goodman during an
engineering night on 2017.28 UT and as described in Section
\ref{subsec:spec}.  While undulations exist in the spectrum, they
correlate with the internal quartz lamp used for flat fielding such
that this object is of spectral type DC as shown in Figure
\ref{fig:newspec}.  This object is similar to WD 0000$-$345 in that a
pure-H model better fits the SED over a pure-He model, yet no Balmer
features are seen in the spectrum.  Thus, we use a pure-He atmospheric
model to characterize this object.

{\bf WD 1242$-$105} was first classified as a hot subdwarf (sdB) by
\citet{1988SAAOC..12....1K} and later identified as a DA WD by
\citet{2003ApJ...586L..95V} who estimated a photometric distance of 18
$\pm$ 4 pc.  It was then placed on the parallax program to confirm
proximity.  With a trigonometric parallax distance of
41.1$\left._{-2.33}^{+2.63}\right.$ pc, we suspected this object to be
an unresolved double star.

\begin{figure}[h!]
\centering
\includegraphics[angle=0, trim={3.6cm 2cm 8.5cm 6.4cm},
  width=0.37\textwidth]{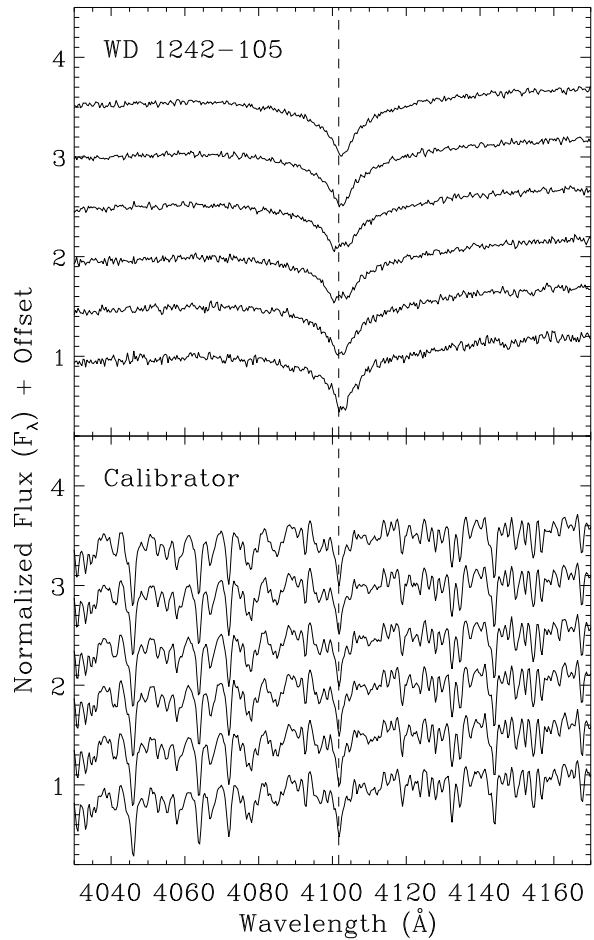}
\caption{SOAR $+$ Goodman stacked spectra for WD 1242$-$105 ({\it top
  panel}) and the calibrator star ({\it bottom panel}).  The dashed
  vertical line denotes the location of H$\beta$.  These spectra cover
  roughly half of the orbital period of WD 1242$-$105.}
\label{fig:wd1242_soarspec}
\end{figure}

During a spectroscopic campaign using the SOAR telescope with the
Goodman spectrograph briefly described in Section \ref{subsec:spec},
we were assisting with the commissioning of a 2100 lines-per-mm VPH
grating and chose to observe this object as a test case.  The
observations were taken on 2011 July 22 and a slit width of
0$\farcs$84 was used to optimize spectral resolution and throughput.
Spectral coverage was 3700$-$4400 \AA~with a spectral resolution of
0.8 \AA~pixel$^{-1}$.  The long slit was oriented so that a bright
calibrator (2MASS J12445203-1049037) was also in the slit to serve as
a check for any wavelength calibration systematic uncertainties
(though it was not known a priori if this calibrator was radial
velocity stable).  The targets were observed for $\sim$2 hours, taking
repeated exposures of 150 seconds each.  After every seven exposures,
arc lamps were taken to ensure accurate wavelength calibration.
Spectroscopic reductions were performed using standard IRAF routines
in the {\it Specred} package.  Once wavelength calibration was
performed on each spectrum (using the arc lamp taken closest in time),
each block of seven spectra was stacked to increase S/N, thus resulting
in a cadence of roughly 25 minutes.  For consistency, the calibrator
spectra were stacked in an identical manner.  See Figure
\ref{fig:wd1242_soarspec}.

\begin{figure}[t!]
\centering
\includegraphics[angle=0, trim={0.2cm 0 11.5cm 22.7cm},
  width=0.45\textwidth]{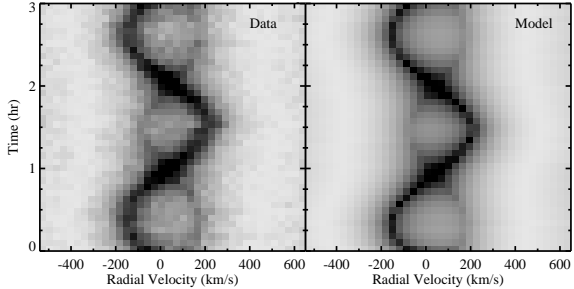}
\caption{Trailed spectrogram of the Gemini-South $+$ GMOS data for
  WD 1242$-$105 around the H$\alpha$ region ({\it left panel}) as well
  as a noiseless model spectrogram using our derived radial velocity
  solution ({\it right panel}).}
\label{fig:wd1242_sgram}
\end{figure}

Once it was evident that the system was a double-lined double
degenerate with rapidly changing relative velocities, we requested the
use of the Gemini Multi-Object Spectrograph-South (GMOS-S) to better
characterize the orbital period.  Spectroscopic data were taken on
2012.31 UT using the R831\_G5323 grating, centered at 6720 \AA, with a
slit width of 0$\farcs$25.  We opted for a central wavelength somewhat
distant from the targeted H$\alpha$ absorption line at 6563
\AA~because the GMOS-S central detector suffered from a bad column
just blueward of center (since been replaced).  With a full wavelength
coverage of 5540$-$7640 \AA~and a $\sim$700 \AA~span within each
detector, the H$\alpha$ region was amply sampled within the central
detector (to avoid the gaps between detectors).  The observing
sequence consisted of a target acquisition on the slit followed by a
block of five spectroscopic observations, each of 300 seconds
integration.  This block of five exposures, followed by an arc lamp,
was repeated four times.  Then, a re-acquisition was performed to
ensure the target was centered in the narrow slit and the observing
sequence was repeated.  The observations span nearly four continuous
hours and resulted in 40 science spectra.  Data were reduced using the
external IRAF package {\it Gemini} Version 1.11 and, in particular,
the suite of routines in the {\it GMOS} subpackage.  Science spectra
were wavelength calibrated with the arc lamp taken nearest in time to
the science spectra.

\begin{figure}[t!]
\centering \includegraphics[angle=0, trim={0.2cm 0 12.5cm 21cm},
  width=0.45\textwidth]{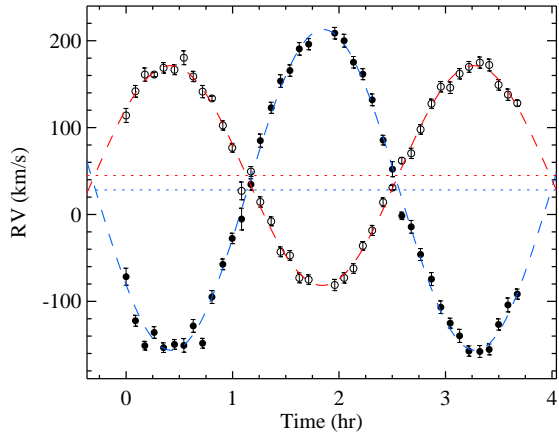}
\caption{Component radial velocity curves for WD 1242$-$105 as a
  function of time.  Open circles and filled circles correspond to
  measures for the more massive component and less massive component,
  respectively.  The horizontal dashed lines represent the $\gamma$
  values with the difference between the two being a result of
  differing gravitational redshifts.  Supplemental data file of these
  radial velocity measurements is available in the online journal.}
\label{fig:wd1242_rvs}
\end{figure}

Spectral line fitting was performed as follows.  The H$\alpha$
absorption lines were fit at maximum separation using pseudo--Gaussian
profiles to determine individual profile shapes (line depths and
widths).  The profile shape was then fixed and applied to all other
spectra with the only free parameter being the pseudo--Gaussian
centroid.  Synthetic H$\alpha$ spectra were then created using these
best-fitting models, with the same resolution, to generate the
noiseless model trailed spectrogram and compared to the data as shown
in Figure \ref{fig:wd1242_sgram}.

These data, once fitted, revealed a double degenerate with a 2.85-hour
period.  The component radial velocity curves as a function of time
are displayed in Figure \ref{fig:wd1242_rvs}.  The differences in the
apparent systemic velocities for each component are because of
differences in component gravitational redshifts.  The vital
parameters for this system are summarized in Table
\ref{tab:wd1242_vitals} and agree very well with those determined by
\citet{2015AJ....149..176D}, as does our parallax determination
compared to those authors' trigonometric parallax.

{\bf WD 1314$-$153} was shown to be, most likely, a young halo WD by
\citet{2012MNRAS.425.1394K} based on kinematics that include radial
velocity measurements as well as tangential motions assuming a
distance of 58 pc.  Our trigonometric parallax distance of 58.1 pc is
entirely consistent with their distance estimate and the space motions
are confirmed.

{\bf WD 1339$-$340} was shown to be a strong halo candidate with a
nearly polar Galactic orbital motion by \citet{2005ApJ...633L.121L}.
They estimate a distance of 18 pc for their analysis, entirely
consistent with our trigonometric parallax distance of 21.0 pc.  With
a \vtan $=$ 255.0 \kmsec, this object has the largest tangential
velocity of the 25 pc WD sample.  Thus, space motions and strong halo
candidacy are confirmed.

\begin{figure}[t!]
\centering
\includegraphics[angle=0, trim={3cm 1.5cm 8cm 11cm},
  width=0.40\textwidth]{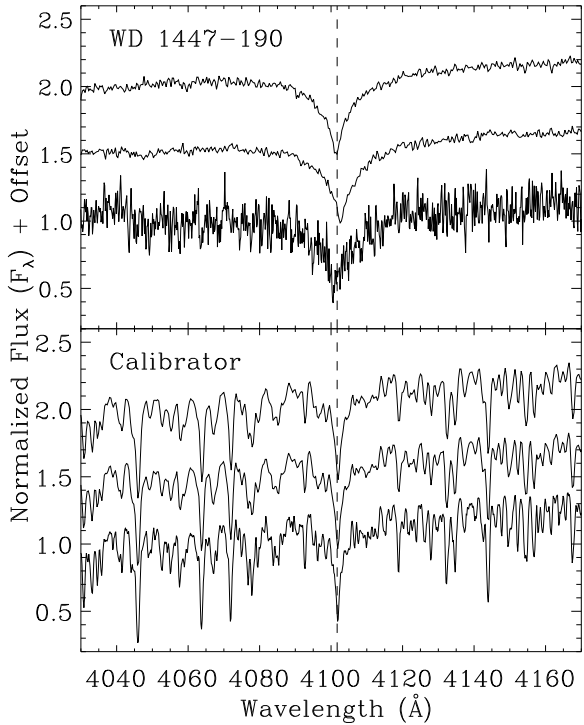}
\caption{SOAR $+$ Goodman stacked spectra for WD 1447$-$190 ({\it top
    panel}) and the calibrator star ({\it bottom panel}).  The dashed
  vertical line denotes the location of H$\beta$.}
\label{fig:wd1447_soarspec}
\end{figure}

{\bf WD 1447$-$190} was first spectroscopically identified as a DA WD
by \citet{2006ApJ...643..402K}.  The photometric distance estimate of
29.1 $\pm$ 4.9 pc \citep{2007AJ....134..252S} is significantly
discrepant from its trigonometric parallax distance of 47.4 $\pm$ 2.0
pc.  Thus, we expected this system to be an unresolved double
degenerate and it served as a second target for the commissioning of
the SOAR $+$ Goodman spectrograph using the 2100 lines-per-mm VPH
grating.  It was observed on three separate nights; 19 July 2011, 22
July 2011, and 6 September 2011.  The instrument setup is similar to
that used for the observations of WD 1242$-$105 and has the same
wavelength coverage.  The first two nights of observations used a
0$\farcs$84 slit width with the slit oriented such that a brighter
star, 2MASS J14500516-1912509, was also included in the slit to act as
a calibrator.  The observing sequence for the first two nights
consisted of four 300 second integrations followed by an arc lamp as a
single block of exposures.  This sequence was repeated eight, and
seven times, respectively, for the first two nights.  The data showed
no short-period variations over the course of each night so all data
for a night were stacked into a single spectrum (see Figure
\ref{fig:wd1447_soarspec}, top two spectra in each panel).  The third
night's data set was smaller as these observations were taken between
core commissioning tasks. The instrument setup consisted of a
0$\farcs$46 wide slit oriented to include the same calibrator star.
Pairs of 600 second exposures were taken followed by an arc lamp.
This sequence was repeated twice.  All data from this night were
stacked into a single spectrum though are noisier than the previous
two nights' spectra.

A clear single-lined radial velocity variation is seen in the WD and
is shown in Figure \ref{fig:wd1447_soarspec}.  The period is
unconstrained with these data but the multiplicity is confirmed.  We
hope to obtain spectroscopic follow-up and better characterize the
orbital parameters of this system.

\begin{figure}[t!]
\centering
\includegraphics[angle=0, trim={1cm 1cm 0 1cm},
  width=0.48\textwidth]{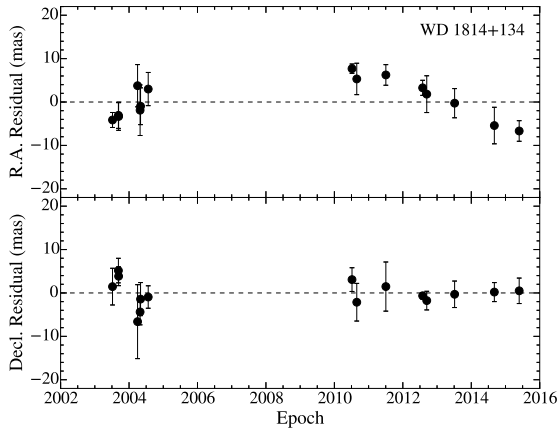}
\caption{CTIOPI nightly mean astrometric residuals of WD 1814$+$134,
  once parallax and proper motion are removed, that show a long-period
  astrometric perturbation.}
\label{fig:wd1814ast}
\end{figure}

{\bf WD 1814$+$134} was first discovered by the SUPERBLINK survey as a
high proper motion star \citep[$\mu =$ 1$\farcs$207 yr$^{-1}$,
][]{2002AJ....124.1190L} and later spectroscopically classified as a
DA10 \citep{2003AJ....125.1598L}.  A trigonometric parallax of 70.3
$\pm$ 1.2 mas was measured by \citet{2009AJ....137.4109L} from data
taken between 2005.48 and 2007.72.  The NOFS trigonometric parallax
(presented here) of 66.05 $\pm$ 0.28 mas was measured from data taken
between 2008.29 and 2012.47 and only marginally agrees with the
previous trigonometric parallax at the 3-$\sigma$ level.  A third
trigonometric parallax of 68.25 $\pm$ 0.96 mas was measured by CTIOPI
(also presented here) and includes data taken between 2003.52 and
2015.40, albeit with a gap from 2005 to 2010 because of the different
$V$ filter used (see Section \ref{subsubsubsec:cracked}).  The CTIOPI
dataset shows a clear, long-period astrometric perturbation, largely
in the right ascension axis, though notable in the declination axis at
the earliest epochs (see Figure \ref{fig:wd1814ast}).  This
perturbation is likely the cause for the discrepant parallax
determinations from data acquired during different phases of the
orbit.  This target remains on the CTIOPI program, and has been
re-added to the NOFS astrometric program to better characterize the
perturbation.  A weighted mean parallax using all three measures has
been adopted for the analysis.  Given that the trigonometric distance
is entirely consistent with atmospheric modeled distance estimates for
a single WD \citep[e.g.,][]{2003AJ....125.1598L, 2007AJ....134..252S},
and there is no noticeable near-IR excess, the companion is likely to
be very low-luminosity.

{\bf WD 2028$-$171} is a newly discovered nearby WD that was targeted
for spectroscopic followup based on a trawl of the NLTT Catalog.  As
discussed in Section \ref{subsec:spec}, we obtained an optical
spectrum of this object and find it to show Ca \textsc{II} H \& K
lines (DAZ) indicative of metal pollution by recent/ongoing accretion.
Follow-up observations are necessary to better characterize this
candidate remnant planetary system.

\begin{figure}[t!]
\centering
\begin{minipage}[b]{0.4\textwidth}
  \centering
  \includegraphics[trim={3.4cm 1cm 3.4cm 1cm},
    width=0.95\textwidth]{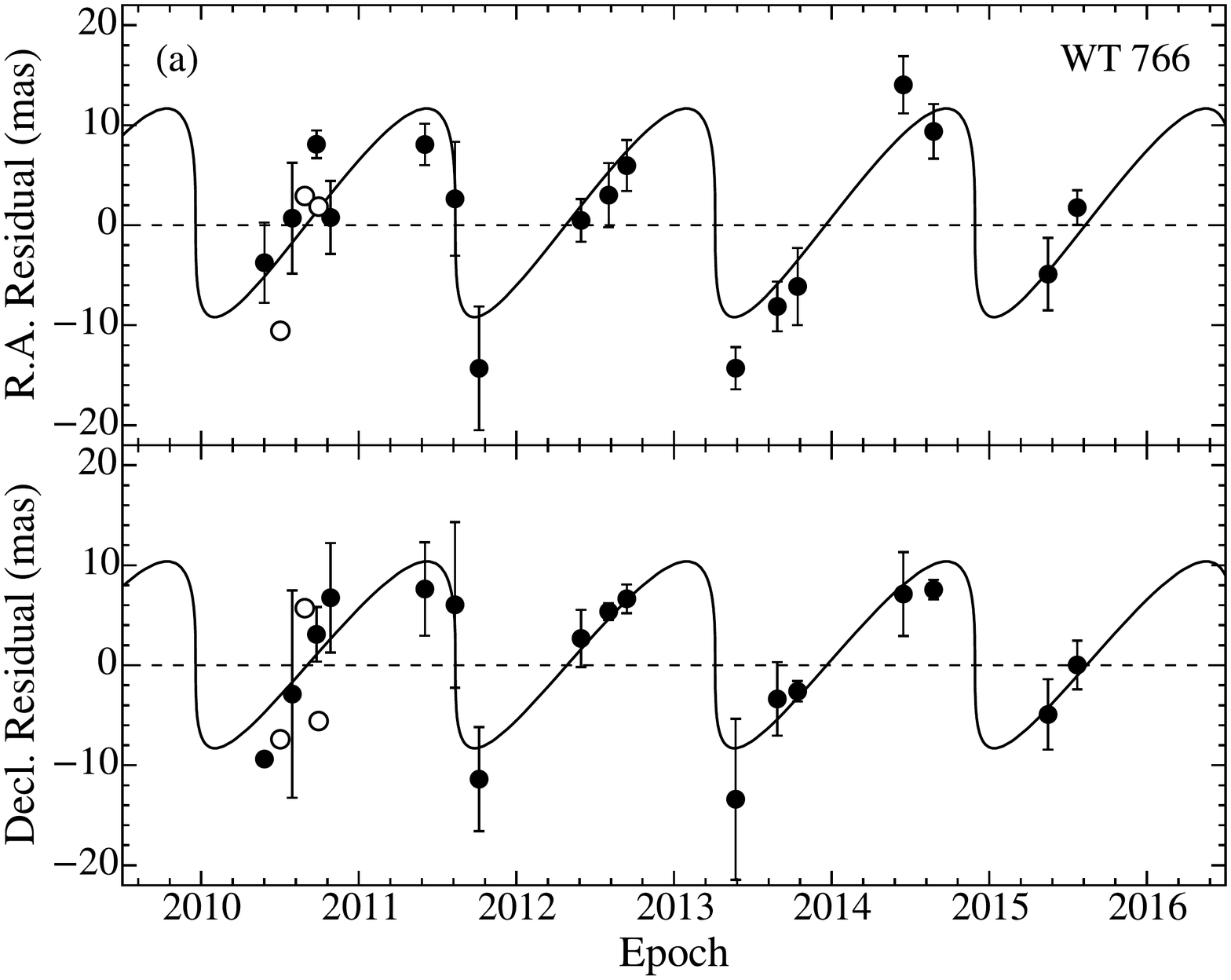}
\end{minipage}
\begin{minipage}[b]{0.4\textwidth}
  \centering
  \includegraphics[trim={3.4cm 1cm 3.4cm 1cm},
    width=0.95\textwidth]{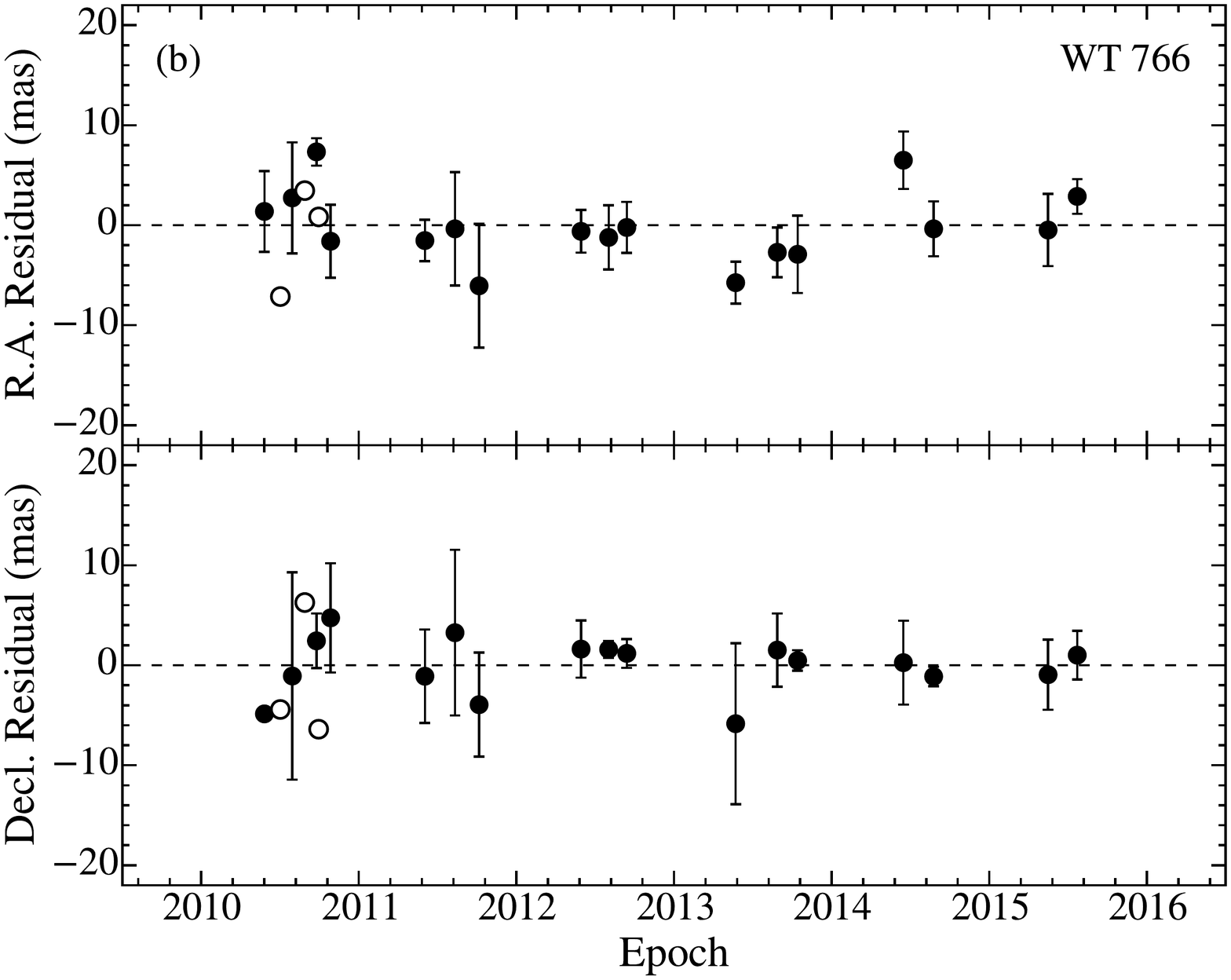}
\end{minipage}
\caption{Nightly mean astrometric residual plots for WT 766, a common
  proper motion companion to WD 2057$-$493.  The upper panel ({\it a})
  shows the raw residuals and the best-fit orbital solution while the
  lower panel ({\it b}) shows the residuals once the orbital solution
  is applied.  In both plots, the open circles represent data for
  which only one frame was taken on that night.  The astrometric
  solution presented in Table \ref{tab:ctioast} for WT 766 accounts
  for the orbital solution.}
\label{fig:wt766_orbit}
\notetoeditor{Please orient these two plots vertically with (a) above
  and (b) below.}
\end{figure}

{\bf WD 2057$-$493} is a newly discovered nearby WD using the
SUPERBLINK database and discussed in Section \ref{subsec:spec}.  It is
also a common proper motion companion to red dwarf WT 766 ($\rho =$
64$\farcs$3 at P.A. 340.3$^\circ$, epoch $=$ 2015.55951 -- measured
from CTIOPI astrometry frames) for which an independent trigonometric
parallax was measured via CTIOPI.  WT 766 shows a clear astrometric
perturbation in the residuals consistent with a period of less than
two years.  The residuals were fit to an orbital model, and while most
orbital parameters (e.g., eccentricity, semi-major axis) are poorly
constrained with an orbital inclination near 90$^\circ$, the period
was well determined to be 1.648 $\pm$ 0.018 years.  The orbital fit is
plotted over the residuals in Figure \ref{fig:wt766_orbit} and removed
from the astrometric analysis to enable a refined trigonometric
parallax.  The parallax determinations for WT 766 and WD 2057$-$493
agree very well, and thus this is a new triple system within 15 pc.

{\bf WD 2159$-$754} is flagged as an ultramassive WD by
\citet{2007ApJ...654..499K} with a mass of 1.17 \msun.  They assume a
distance of 14 pc for their analysis.  Our trigonometric parallax
distance of 19.9 pc results in a slightly lower mass (though still
massive) of 0.97 \msun.

{\bf WD 2226$-$754AB} is a widely separated double degenerate pair
consisting of two cool, featureless DC WDs, originally discovered by
\citet{2002ApJ...565..539S}.  While the tangential velocity of this
pair at $\sim$135 \kmsec~is only marginally consistent with halo
kinematics, the calculated space motion (assuming zero tangential
velocity as none has been measured) is more convincing with $V \sim
-$112 \kmsec~(positive in the direction of Galactic rotation).
Coupled with the WD cooling ages and masses for each component (A:
$\sim$6.5 Gyr and 0.50 \msun; B: $\sim$7.4 Gyr and 0.51 \msun), we
nominally classify this system as a strong halo candidate.

{\bf WD 2307$-$691} is a common proper motion companion, found at
J2000 coordinates 23:10:22.96 -68:50:20.2 (epoch 2011.6966) to HIP
114416 (GJ 1280, LTT 9387), yet has not been characterized in the
literature.  It was brought to the authors' attention by Brian Skiff
(private communication) as a blue companion to a {\it Hipparcos} star
within 25 pc.  We note that the recent work of
\citet{2016MNRAS.462.2295H} has also identified this object as a WD
based on its colors but does not have a spectrum and instead assumes
it to be a DA WD.  We obtained an identification spectrum with the
SOAR $+$ Goodman spectrograph as described in Section
\ref{subsec:spec}.  The spectrum, shown in Figure \ref{fig:newspec},
indicates the object is a relatively cool DB WD and its companion
distance via {\it Hipparcos} \& TGAS of 21.1 $\pm$ 0.1 pc is adopted
for this WD making it a new member of the 25 pc WD sample.

{\bf WD 2326$+$049}, better known as G29$-$38, is a variable ZZ Ceti
WD \citep{1974AFZ...10..117S} with significant IR excess
\citep{1987Natur.330..138Z} and one of the first WDs found with metal
pollution \citep{1997A&A...320L..57K}.  Thus, the spectral
classification is DAZV.  The previous trigonometric parallax from YPC
of 73.4 $\pm$ 4.0 mas is significantly larger than our measure of
56.83 $\pm$ 0.39 mas.  This distance discrepancy will have an impact
on the implied surface gravity and, thus, on settling times for the
accreted metals.  The abundance analysis performed by
\citet{2014ApJ...783...79X} notes that the implied surface gravity of
8.4 dex from the previous parallax measurement produced inconsistent
model spectroscopic line profiles from those observed, suggesting that
a lower surface gravity is likely.  Indeed, our updated parallax
suggests a \logg $=$ 8.00 $\pm$ 0.03.  Because the
IR excess affects the near-IR $HK_s$ photometry, these measures have
been excluded from the model fit.


\section{Discussion}
\label{sec:disc}
\subsection{25 pc White Dwarf Sample}
\label{subsec:25pc}

\begin{figure}[t!]
\centering
\includegraphics[angle=0, trim={0.5cm 0.5cm 0 0.7cm},
  width=0.46\textwidth]{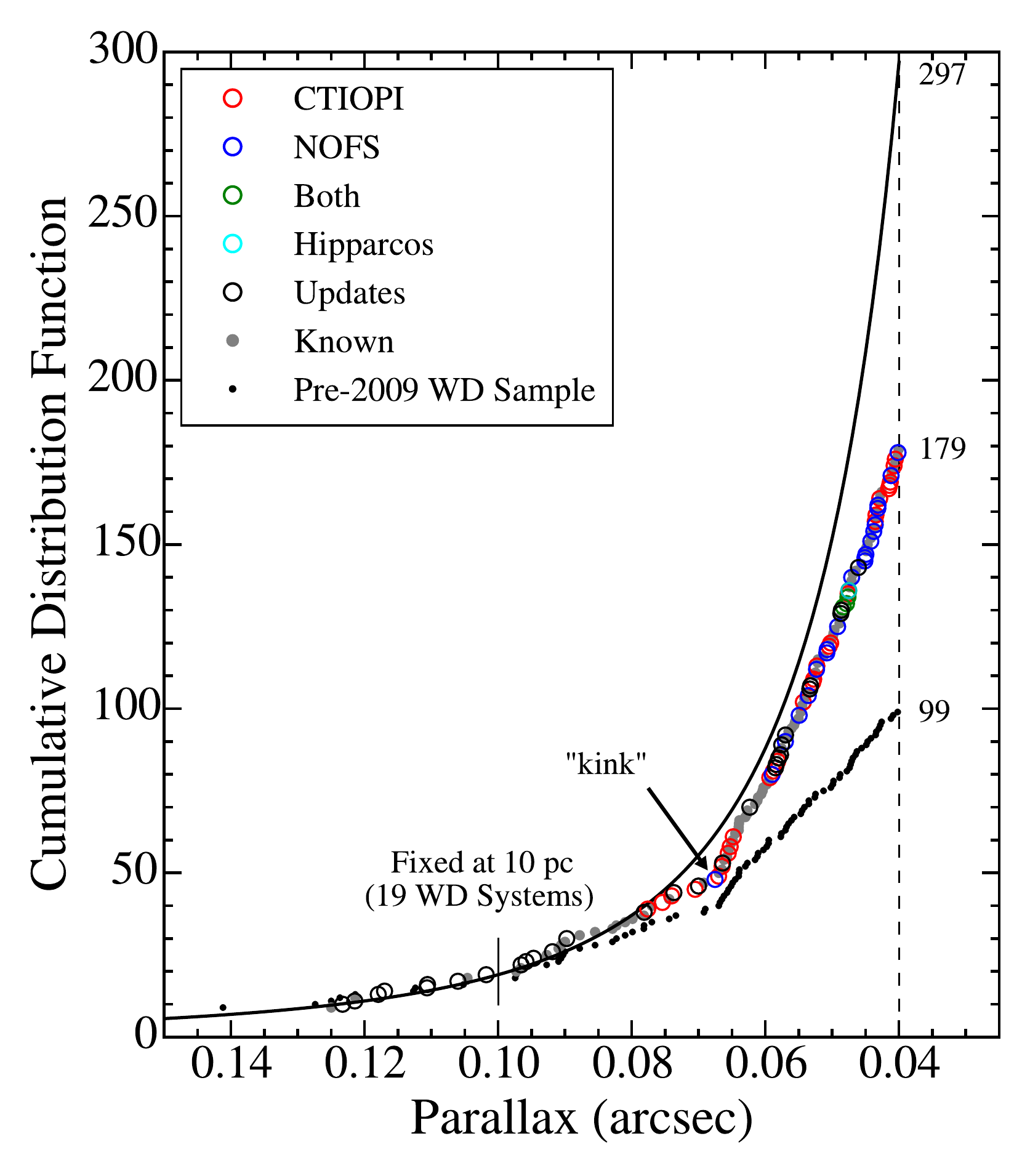}
\caption{Cumulative distribution plot for the 25 pc WD sample.  New
  additions presented in this work are highlighted by sub-sample as
  shown in the legend.  The black points represent the 25 pc WD sample
  prior to 2009 and serve as an indicator of the progress made since.
  The numbers to the right of the curves (99, 179 and 297) represent
  the number of systems at 25 pc prior to 2009, now, and that expected
  at 25 pc, respectively, indicating $\sim$40\% incompleteness
  currently.  The ``kink'' described in Section \ref{subsec:25pc} is
  identified with an arrow.}
\label{fig:cumdist}
\end{figure}

Prior to the CTIOPI effort to obtain a volume-limited sample of WDs
within 25 pc, whose first WD results were published in 2009
\citep{2009AJ....137.4547S}, a total of 112 systems, of which 99 had
robust trigonometric parallax determinations (i.e., better than 10\%),
were known.  The contributions to the 25 pc sample, both prior to 2009
and since, are tabulated in Table \ref{tab:contributions}.  Prior to
this work, the 25 pc WD sample consisted of 137 systems, of which 126
systems have robust trigonometric parallax determinations.  The
trigonometric parallaxes presented in this work consist of 23 new
systems from CTIOPI, 19 new systems from NOFS, one new system measured
by both programs, as well as seven of the eleven systems with poor
previous trigonometric parallax determinations, i.e., worse than 10\%.
Two additional systems, one new (WD 0148$+$641) and one with a poor
previous parallax determination (WD 2117$+$539), have parallax
determinations from NOFS and are also included in the recently
published TGAS catalog with measurements confirming 25 pc sample
membership.  Finally, we spectroscopically confirm that WD 2307$-$691
is a WD companion to HIP 114416, with a {\it Hipparcos}$+$TGAS
trigonometric parallax distance of 21.0 $\pm$ 0.1 pc.  The final
number of members added by this work is 53 systems, thereby increasing
the sample completeness by 42\% to 179 systems.  The measured local WD
density is then 2.7 $\times$ 10$^{-3}$ pc$^{-3}$ based on this sample.
This value is considerably lower than that found by
\citet{2016MNRAS.462.2295H} of 4.8 $\times$ 10$^{-3}$ pc$^{-3}$,
because those authors include the expected incompleteness based on a
complete 13 pc sample.  As shown in Figure \ref{fig:cumdist}, assuming
a constant density and that all WD systems within 10 pc are known, we
expect this sample to be incomplete at the $\sim$40\% level even after
the addition of the new members presented here.  Accounting for this
expected incompleteness, we arrive at an expected local WD density of
4.5 $\times$ 10$^{-3}$ pc$^{-3}$, very similar to that found by
\citet{2016MNRAS.462.2295H}.  Another curiosity seen in the figure is
the ``kink'' in observed systems around 15 pc, thus implying a dearth
of WD systems at that distance.  While a few new systems were found at
and within that distance, pure incompleteness from single field WDs is
not likely a significant factor.  There is the possibility that new WD
systems will be identified as companions to bright main-sequence
primaries (the so-called ``Sirius-like'' systems) that may reduce the
dearth.

{\it Gaia} is a cornerstone mission in the science programme of the
European Space Agency (ESA) that will deliver unprecedented
astrometric accuracy for one billion stars in the Milky Way.  A
description of the mission is given by \citet{2016A&A...595A...1G}.
While {\it Gaia} will effectively complete the 25 pc WD sample and
beyond \citep[][estimates effective WD completeness by {\it Gaia} out
  to $\sim$55 pc]{2015ASPC..493..455S}, the DR1 released in 2016
\citep{2016A&A...595A...2G} only includes trigonometric parallaxes for
      {\it Hipparcos} and {\it Tycho} stars and thus, did not contain
      new sample members beyond the two mentioned above.

\subsection{Sky Distribution}
\label{subsec:skydist}

\begin{figure*}[t!]
\centering
\includegraphics[angle=0, trim={3cm 5cm 3cm 5cm},
  width=0.95\textwidth]{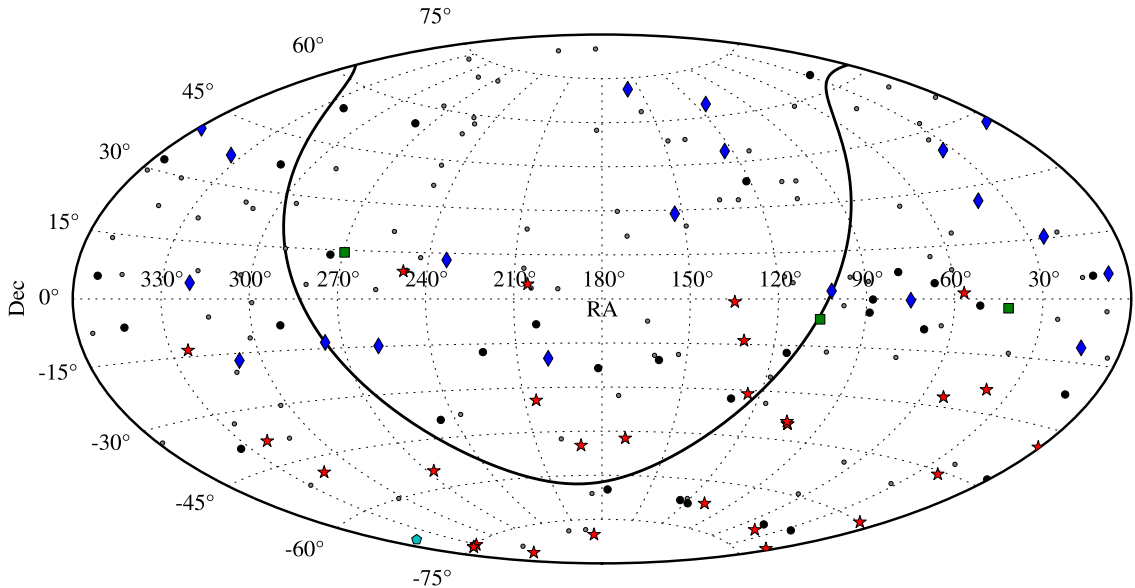}
\caption{Aitoff projection of the 25 pc WD sample.  The curve
  represents the Galactic Plane.  Previously known WD systems ({\it
    small gray circles}) and updates ({\it large filled circles}) are
  shown.  New additions from CTIOPI ({\it red stars}), NOFS ({\it blue
    diamonds}), both programs ({\it green squares}), as well as the
  lone {\it Hipparcos} companion ({\it cyan pentagon}) are also shown.
  Note that the two systems measured by the {\it Gaia} TGAS survey are
  included in the NOFS sample for which parallaxes were confirmed.}
\label{fig:aitoff}
\notetoeditor{If this figure can span two columns, that would be ideal.}
\end{figure*}

Traditionally, the southern hemisphere has been undersampled relative
to the north simply because there are more observing assets in the
north.  In particular, as nearby stars tend to be identified via
proper motion surveys, it is only recently that proper motion surveys
in the south \citep[e.g.,][]{2005AAS...20715001L,2011AJ....142...92B}
have pushed to sufficiently low proper motions to better identify
nearby, slow movers.  With the samples presented here, we have
effectively removed the bias towards northern targets (see Table
\ref{tab:skydist}) as there are now almost identical numbers, 90 and
89 in the northern and southern hemispheres, respectively.  An Aitoff
projection of the 25 pc WD sky distribution is shown in Figure
\ref{fig:aitoff}.

\section{Conclusions}
\label{sec:conclusions}

With respect to individual systems presented here, we find four that
are very strong halo candidates (WD 1237$-$230, \vtan $=$ 202.4
\kmsec; WD 1314$-$153, \vtan $=$ 191.7 \kmsec; WD 1339$-$340, \vtan
$=$ 255.0 \kmsec; and WD 2226$-$754AB, \vtan $=$ 135.3 \kmsec) based
on kinematics alone and they often have other supporting evidence
collected by other researchers.  Two of these systems are mere
interlopers that happen to be within the 25 pc volume at present (WD
1339$-$340 and WD 2226$-$754AB).  As discussed in Section
\ref{subsec:comments}, WD 1237$-$230 is an unresolved double
degenerate so our mass estimate is incorrect.  For the three other
halo candidates, the masses are entirely consistent with old total
ages, i.e., when the main-sequence progenitor lifetimes are taken into
account \citep[see][their Figure 9]{2005ApJ...625..838B} expected for
halo membership.

We find two very cool WDs that display CIA such that they appear
relatively blue (WD 0222$-$291 and WD 0851$-$246).  Additionally,
near-IR photometry collected during this effort shows WD 0851$-$246 to
have excess in $JHK_S$ suggesting a red unseen companion to the WD which
is, itself, a common proper motion companion to a widely separated
cool, old subdwarf.

We spectroscopically identify a metal-rich DAZ WD (WD 2028$-$171) and
confirm it to be within the 25 pc volume.  Follow-up observations will
enable a thorough analysis of this remnant planetary system signature.

Our sample contains at least three systems that appear overluminous
(WD 1237$-$230, WD 1242$-$105, and WD 1447$-$190) and, thus, are
likely unresolved multiples.  We conducted follow-up spectroscopic
observations for two (WD 1242$-$105 and WD 1447$-$190) and find that
WD 1242$-$105 is a short-period double degenerate, confirming the
independent work of \cite{2015AJ....149..176D}, and that WD 1447$-$190
is a single-lined spectroscopic binary with a radial velocity
variation over a few days, but have not constrained the orbit.  Radial
velocity work by \cite{2012MNRAS.425.1394K} confirms that WD 1237$-$230
is a single-lined binary.  Two WDs are marginally overluminous (WD
0233$-$242 and WD 1817$-$598); however, they are both
spectroscopically featureless DC WDs that prohibit radial velocity
analyses for confirmation.  We see no astrometric perturbations for
either of these two systems.

Two of our objects do show astrometric perturbations.  WD 1814$+$134
shows a long-period residual that has gone unrecognized in short
baseline astrometric datasets but is visible from the $\sim$13-year
baseline of the CTIOPI astrometric data.  WD 2057$-$493 is a newly
identified nearby WD that we spectroscopically identified in this work
and has a main-sequence common proper motion companion (likely an
early M-dwarf though no spectral confirmation was acquired).
Astrometric data were collected on both members of the system and the
main-sequence star shows a clear astrometric perturbation with a
period of 1.648 years.  The astrometric dataset encompasses $\sim$3
full periods and, given that the astrometric solution for the WD
companion used an identical reference field and shows flat residuals,
the perturbation detection is robust.  

Of the 107 systems whose trigonometric parallaxes are presented here,
50 are new members of the 25 pc WD sample.  In addition, two systems
(WD 0148$+$641 and WD 2117$+$539) have recently determined
trigonometric parallaxes from the TGAS catalog with which our
trigonometric parallaxes agree and confirm sample membership.
Finally, we spectroscopically confirmed a newly identified WD
companion (WD 2307$-$691) to {\it Hipparcos} star HIP 114416 with a
trigonometric parallax within 25 pc that we adopt for the WD
companion.  Thus, a total of 53 new 25 pc WD systems have been added
to a sample of 126 prior systems with robust distance determinations
resulting in a 42\% increase.  Between the CTIOPI and NOFS parallax
programs overall, a total of 70 new WD systems have been added to the
25 pc WD sample -- a 64\% increase to the sample.  It is expected that
{\it Gaia} will largely complete this sample upon release of its final
catalog (if not in one of the early data releases), save for, perhaps,
in regions near the Galactic plane where crowding will cause greater
incompleteness.  This 25 pc sample represents the pre-{\it Gaia}
collection of knowledge for the sample and will serve to clearly
demonstrate the (most likely) significant contribution that {\it Gaia}
will make in this arena.

\acknowledgments 

We would like to thank the referee for a number of comments and
suggestions that have clarified several points.  We thank the members
of the SMARTS Consortium, who enable the operations of the small
telescopes at CTIO, as well as the observer support at CTIO,
specifically Edgardo Cosgrove, Arturo Gomez, Alberto Miranda, and
Joselino Vasquez.  We thank Nicole van der Bliek, Andrea Kunder, and
Ron Probst for their assistance with the NEWFIRM dataset.  We thank
the observer support at NOFS, specifically Michael DiVittorio, Fred
Harris, Albert Rhodes, and Michael Schultheis.  Additionally, we thank
the individuals that have assisted with the NOFS parallax program's
observations over the many years of operation, including Blaise
Canzian, Harry Guetter, Stephen Levine, Christian Luginbuhl, Jeff
Munn, and Trudy Tilleman.

The RECONS team wishes to thank the NSF (grants AST 05$-$07711, AST
09$-$08402, and AST 14$-$12026) and GSU for their continued support in
our study of nearby stars.  P.~B.~ and P.~D.~wish to acknowledge the
support of NSERC Canada as well as the the Fund FRQNT (Qu\'ebec).

Based on observations (GS-2012A-Q-36) obtained at the Gemini
Observatory, acquired through the Gemini Science Archive and processed
using the Gemini IRAF package, which is operated by the Association of
Universities for Research in Astronomy, Inc., under a cooperative
agreement with the NSF on behalf of the Gemini partnership: the
National Science Foundation (United States), the National Research
Council (Canada), CONICYT (Chile), Ministerio de Ciencia,
Tecnolog\'{i}a e Innovaci\'{o}n Productiva (Argentina), and
Minist\'{e}rio da Ci\^{e}ncia, Tecnologia e Inova\c{c}\~{a}o (Brazil).

Based on observations obtained at the Southern Astrophysical Research
(SOAR) telescope, which is a joint project of the Minist\'{e}rio da
Ci\^{e}ncia, Tecnologia, e Inova\c{c}\~{a}o (MCTI) da Rep\'{u}blica
Federativa do Brasil, the U.S. National Optical Astronomy Observatory
(NOAO), the University of North Carolina at Chapel Hill (UNC), and
Michigan State University (MSU).

This publication makes use of data products from the Two Micron All
Sky Survey, which is a joint project of the University of
Massachusetts and the Infrared Processing and Analysis
Center/California Institute of Technology, funded by the National
Aeronautics and Space Administration and the National Science
Foundation.

This work has made use of data from the European Space Agency (ESA)
mission {\it Gaia} (\url{http://www.cosmos.esa.int/gaia}), processed
by the {\it Gaia} Data Processing and Analysis Consortium (DPAC,
\url{http://www.cosmos.esa.int/web/gaia/dpac/consortium}). Funding for
the DPAC has been provided by national institutions, in particular the
institutions participating in the {\it Gaia} Multilateral Agreement.

Funding for SDSS-III has been provided by the Alfred P. Sloan
Foundation, the Participating Institutions, the National Science
Foundation, and the U.S. Department of Energy Office of Science. The
SDSS-III web site is http://www.sdss3.org/.

SDSS-III is managed by the Astrophysical Research Consortium for the
Participating Institutions of the SDSS-III Collaboration including the
University of Arizona, the Brazilian Participation Group, Brookhaven
National Laboratory, University of Cambridge, Carnegie Mellon
University, University of Florida, the French Participation Group, the
German Participation Group, Harvard University, the Instituto de
Astrofisica de Canarias, the Michigan State/Notre Dame/JINA
Participation Group, Johns Hopkins University, Lawrence Berkeley
National Laboratory, Max Planck Institute for Astrophysics, Max Planck
Institute for Extraterrestrial Physics, New Mexico State University,
New York University, Ohio State University, Pennsylvania State
University, University of Portsmouth, Princeton University, the
Spanish Participation Group, University of Tokyo, University of Utah,
Vanderbilt University, University of Virginia, University of
Washington, and Yale University.



\clearpage


\voffset80pt{}
\begin{deluxetable}{lcccccccccccccr@{$\pm$}rr@{$\pm$}rr@{$\pm$}rc}
\rotate
\tabletypesize{\tiny}
\tablewidth{0pt}
\tablecaption{Photometric Results
\label{tab:pht}}
\tablehead{
  \colhead{}                                 &
  \colhead{Alternate}                        &
  \colhead{}                                 &
  \colhead{}                                 &
  \colhead{}                                 &
  \colhead{}                                 &
  \colhead{}                                 &
  \colhead{}                                 &
  \colhead{}                                 &
  \colhead{}                                 &
  \colhead{}                                 &
  \colhead{$\sigma$}                         &
  \colhead{No. of}                           &
  \colhead{No. of}                           &
  \multicolumn{2}{c}{}                       &
  \multicolumn{2}{c}{}                       &
  \multicolumn{2}{c}{}                       &
  \colhead{}                                \\

  \colhead{Name}                             &
  \colhead{Name}                             &
  \colhead{$B_{\rm J}$}                      &
  \colhead{$N_{\rm B}$}                      &
  \colhead{$V_{\rm J}$}                      &
  \colhead{$N_{\rm V}$}                      &
  \colhead{$R_{\rm KC}$}                     &
  \colhead{$N_{\rm R}$}                      &
  \colhead{$I_{\rm KC}$}                     &
  \colhead{$N_{\rm I}$}                      &
  \colhead{$\pi$ Filter}                     &
  \colhead{(mag)}                            &
  \colhead{Nights}                           &
  \colhead{Frames}                           &
  \multicolumn{2}{c}{$J_{\rm 2M}$}           &
  \multicolumn{2}{c}{$H_{\rm 2M}$}           &
  \multicolumn{2}{c}{$K_{\rm S_{\rm 2M}}$}   &
  \colhead{Notes}                           \\

  \colhead{(1)}                              &
  \colhead{(2)}                              &
  \colhead{(3)}                              &
  \colhead{(4)}                              &
  \colhead{(5)}                              &
  \colhead{(6)}                              &
  \colhead{(7)}                              &
  \colhead{(8)}                              &
  \colhead{(9)}                              &
  \colhead{(10)}                             &
  \colhead{(11)}                             &
  \colhead{(12)}                             &
  \colhead{(13)}                             &
  \colhead{(14)}                             &
  \multicolumn{2}{c}{(15)}                   &
  \multicolumn{2}{c}{(16)}                   &
  \multicolumn{2}{c}{(17)}                   &
  \colhead{(18)}}

\startdata
WD2359$-$434    & LHS 1005           & \nodata &\nodata&  12.97  &   3   &  12.82  &   3   &  12.66  &   3   &    R    &  0.007  &   23    &   142   &  12.60  &   0.03   &  12.43  &   0.02   &  12.45  &   0.02   & \tablenotemark{a}   \\ 
WD0000$-$345    & LHS 1008           &  15.42  &   2   &  15.03  &   4   &  14.73  &   4   &  14.43  &   4   &    R    &  0.007  &   14    &    61   &  14.12  &   0.02   &  14.02  &   0.04   &  13.92  &   0.06   &    \\ 
WD0008$+$424    & GD 5               & \nodata &\nodata& \nodata &\nodata& \nodata &\nodata& \nodata &\nodata& \nodata & \nodata & \nodata & \nodata &  14.54  &   0.03   &  14.35  &   0.04   &  14.39  &   0.06   &    \\
WD0011$-$721    & LP 50$-$73         & \nodata &\nodata&  15.17  &   3   &  14.87  &   3   &  14.55  &   3   &    R    &  0.006  &   13    &    66   &  14.21  &   0.03   &  13.97  &   0.04   &  13.91  &   0.05   & \tablenotemark{b}   \\ 
WD0025$+$054    & LSPM J0027$+$0542  &  16.79  &   1   &  16.15  &   1   &  15.76  &   1   &  15.38  &   1   & \nodata & \nodata & \nodata & \nodata &  14.92  &   0.02   &  14.65  &   0.02   &  14.56  &   0.02   & \tablenotemark{c}   \\
WD0034$-$602    & LP 122$-$4         &  14.19  &   2   &  14.08  &   3   &  14.19  &   3   &  14.20  &   3   &    R    &  0.006  &   14    &    59   &  14.37  &   0.04   &  14.55  &   0.06   &  14.52  &   0.09   & \tablenotemark{d}   \\ 
WD0038$-$226    & GJ 2012            &  15.15  &   2   &  14.49  &   5   &  14.08  &   4   &  13.67  &   5   &    R    &  0.006  &   29    &   133   &  13.34  &   0.03   &  13.48  &   0.03   &  13.74  &   0.04   & \tablenotemark{e}   \\ 
WD0046$+$051    & LHS 7              &  12.91  &   2   &  12.38  &   3   &  12.15  &   3   &  11.89  &   3   &    R    &  0.006  &   15    &    77   &  11.69  &   0.02   &  11.57  &   0.02   &  11.50  &   0.03   &    \\
\enddata

\tablenotetext{a}{ Optical photometry values are adopted from
  \citet{2009AJ....137.4547S}.}
\tablenotetext{b}{ Optical photometry values are adopted from
  \citet{2008AJ....136..899S}.}
\tablenotetext{c}{ $JHK_S$ photometry has been transformed from the
  UKIDSS system using the transformations of
  \citet{2009MNRAS.394..675H} to the 2MASS system.}
\tablenotetext{d}{ Optical photometry values are adopted from
  \citet{2007AJ....134..252S}.}
\tablenotetext{e}{ Optical photometry values include additional
  measures than those presented in \citet{2009AJ....137.4547S} and,
  thus, supersede the values presented in that publication.}
\tablenotetext{f}{ Optical photometry values include additional
  measures than those presented in \citet{2007AJ....134..252S} and,
  thus, supersede the values presented in that publication.}
\tablenotetext{g}{ Likely variable at the $\sim$1\%$-$2\% level.}
\tablenotetext{h}{ Optical photometry values include additional
  measures than those presented in \citet{2008AJ....136..899S} and,
  thus, supersede the values presented in that publication.}
\tablenotetext{i}{ Variability analysis is contaminated by a nearby
  source, hence the brackets in Column 12 indicating erroneous
  variability.}
\tablenotetext{j}{ Optical magnitudes for this close binary ($\rho =
  $~6$\farcs$3 at P.A. 226.5$^\circ$, epoch $=$ 2014.91832) are
  uncertain because of the marginal calibrators for PSF photometry in
  the 6$\farcm$8$^2$ FOV.}
\tablenotetext{k}{ $JHK_S$ magnitudes are from NEWFIRM and are on the
  MKO system and, thus, represented with an italic font.}

\end{deluxetable}
\clearpage


\voffset90pt{}
\begin{deluxetable}{lllccrccrr@{$\pm$}rr@{$\pm$}rr@{$\pm$}rr@{$\pm$}rr@{$\pm$}rrl}
\rotate
\tabletypesize{\tiny}
\tablewidth{0pt}
\tablecaption{CTIOPI Astrometric Results
\label{tab:ctioast}}
\tablehead{
  \colhead{}                                 &
  \colhead{}                                 &
  \colhead{}                                 &
  \colhead{}                                 &
  \colhead{}                                 &
  \colhead{}                                 &
  \colhead{}                                 &
  \colhead{}                                 &
  \colhead{}                                 &
  \multicolumn{2}{c}{$\pi$(rel)}             &
  \multicolumn{2}{c}{$\pi$(corr)}            &
  \multicolumn{2}{c}{$\pi$(abs)}             &
  \multicolumn{2}{c}{$\mu$}                  &
  \multicolumn{2}{c}{P.A.}                   &
  \colhead{$V_{\rm tan}$}                    &
  \colhead{}                                \\

  \colhead{Name}                             &
  \colhead{R.A. (J2000.0)}                   &
  \colhead{Decl. (J2000.0)}                  &
  \colhead{Filter}                           &
  \colhead{$N_{\rm sea}$}                    &
  \colhead{$N_{\rm frm}$}                    &
  \colhead{Coverage}                         &
  \colhead{Years}                            &
  \colhead{$N_{\rm ref}$}                    &
  \multicolumn{2}{c}{(mas)}                  &
  \multicolumn{2}{c}{(mas)}                  &
  \multicolumn{2}{c}{(mas)}                  &
  \multicolumn{2}{c}{(mas yr$^{\rm -1}$)}    &
  \multicolumn{2}{c}{(deg)}                  &
  \colhead{(km s$^{\rm -1}$)}                &
  \colhead{Notes}                           \\

  \colhead{(1)}                              &
  \colhead{(2)}                              &
  \colhead{(3)}                              &
  \colhead{(4)}                              &
  \colhead{(5)}                              &
  \colhead{(6)}                              &
  \colhead{(7)}                              &
  \colhead{(8)}                              &
  \colhead{(9)}                              &
  \multicolumn{2}{c}{(10)}                   &
  \multicolumn{2}{c}{(11)}                   &
  \multicolumn{2}{c}{(12)}                   &
  \multicolumn{2}{c}{(13)}                   &
  \multicolumn{2}{c}{(14)}                   &
  \colhead{(15)}                             &
  \colhead{(16)}}

\startdata

WD2359$-$434    & 00 02 10.73 & $-$43 09 55.6 & R & 13s & 142 & 2003.77$-$2015.83 & 12.06 &   7 & 120.12 & 0.88 & 1.25 & 0.19 & 121.37 & 0.90 &  886.8 & 0.2 & 138.8 & 0.03 &   34.6 & \tablenotemark{a}   \\ 
WD0000$-$345    & 00 02 40.10 & $-$34 13 39.5 & R &  6s &  61 & 2010.50$-$2015.84 &  5.34 &   8 &  69.36 & 1.41 & 1.03 & 0.12 &  70.39 & 1.42 &  766.5 & 0.8 & 168.8 & 0.09 &   51.6 & \tablenotemark{b, {\rm c}}   \\ 
WD0011$-$721    & 00 13 49.91 & $-$71 49 54.3 & R &  5s &  66 & 2007.55$-$2011.97 &  4.42 &   8 &  52.15 & 1.10 & 0.62 & 0.03 &  52.77 & 1.10 &  335.7 & 0.7 & 137.2 & 0.23 &   30.1 & \tablenotemark{b}   \\ 
WD0034$-$602    & 00 36 22.32 & $-$59 55 27.5 & R &  6s &  59 & 2004.56$-$2009.92 &  5.36 &  10 &  40.24 & 1.56 & 1.17 & 0.14 &  41.41 & 1.57 &  276.8 & 0.7 &  63.2 & 0.28 &   31.7 & \tablenotemark{b}   \\ 
WD0038$-$226    & 00 41 26.03 & $-$22 21 02.3 & R & 17s & 133 & 1999.64$-$2015.82 & 16.19 &   7 & 110.05 & 0.86 & 1.24 & 0.07 & 111.29 & 0.86 &  608.0 & 0.2 & 232.6 & 0.03 &   25.9 & \tablenotemark{a}   \\ 
WD0046$+$051    & 00 49 09.91 & $+$05 23 19.1 & R &  9s &  77 & 2005.80$-$2015.96 & 10.16 &   6 & 230.87 & 1.56 & 1.25 & 0.09 & 232.12 & 1.56 & 2970.6 & 0.4 & 156.9 & 0.02 &   60.7 &    \\ 
WD0123$-$460    & 01 25 18.04 & $-$45 45 31.2 & R &  8s &  63 & 2005.70$-$2012.94 &  7.24 &   9 &  42.61 & 1.88 & 0.28 & 0.03 &  42.89 & 1.88 &  744.2 & 0.7 & 139.2 & 0.10 &   82.2 & \tablenotemark{b}   \\ 
WD0127$-$311    & 01 29 56.13 & $-$30 55 09.9 & V &  6g &  59 & 1999.71$-$2011.73 & 12.02 &   5 &  18.04 & 1.19 & 1.11 & 0.08 &  19.15 & 1.19 &  120.6 & 0.4 & 187.6 & 0.27 &   29.8 & \tablenotemark{c}   \\
\enddata

\tablenotetext{a}{Astrometric determinations presented here supersede
  those of \citet{2009AJ....137.4547S} because they include additional
  data over $\sim$6 years (see Section \ref{sec:astres}).}
\tablenotetext{b}{New member of the 25 pc WD sample.}
\tablenotetext{c}{Object is discussed in Section
  \ref{subsec:comments}.}
\tablenotetext{d}{Both $oV$ and $nV$ data were used to determine the
  astrometric solution as described in Section
  \ref{subsubsubsec:cracked}.}

\end{deluxetable}
\clearpage


\voffset100pt{}
\begin{deluxetable}{lllccrccrr@{$\pm$}rr@{$\pm$}rr@{$\pm$}rr@{$\pm$}rr@{$\pm$}rrll}
\rotate
\tabletypesize{\tiny}
\tablewidth{0pt}
\tablecaption{NOFS Astrometric Results
\label{tab:nofsast}}
\tablehead{
  \colhead{}                                 &
  \colhead{}                                 &
  \colhead{}                                 &
  \colhead{}                                 &
  \colhead{}                                 &
  \colhead{}                                 &
  \colhead{}                                 &
  \colhead{}                                 &
  \colhead{}                                 &
  \multicolumn{2}{c}{$\pi$(rel)}             &
  \multicolumn{2}{c}{$\pi$(corr)}            &
  \multicolumn{2}{c}{$\pi$(abs)}             &
  \multicolumn{2}{c}{$\mu$}                  &
  \multicolumn{2}{c}{P.A.}                   &
  \colhead{$V_{\rm tan}$}                    &
  \colhead{}                                &
  \colhead{}                                \\

  \colhead{Name}                             &
  \colhead{R.A. (J2000.0)}                   &
  \colhead{Decl. (J2000.0)}                  &
  \colhead{Filter}                           &
  \colhead{$N_{\rm ngt}$}                    &
  \colhead{$N_{\rm frm}$}                    &
  \colhead{Coverage}                         &
  \colhead{Years}                            &
  \colhead{$N_{\rm ref}$}                    &
  \multicolumn{2}{c}{(mas)}                  &
  \multicolumn{2}{c}{(mas)}                  &
  \multicolumn{2}{c}{(mas)}                  &
  \multicolumn{2}{c}{(mas yr$^{\rm -1}$)}    &
  \multicolumn{2}{c}{(deg)}                  &
  \colhead{(km s$^{\rm -1}$)}                &
  \colhead{CCD}                             &
  \colhead{Notes}                           \\

  \colhead{(1)}                              &
  \colhead{(2)}                              &
  \colhead{(3)}                              &
  \colhead{(4)}                              &
  \colhead{(5)}                              &
  \colhead{(6)}                              &
  \colhead{(7)}                              &
  \colhead{(8)}                              &
  \colhead{(9)}                              &
  \multicolumn{2}{c}{(10)}                   &
  \multicolumn{2}{c}{(11)}                   &
  \multicolumn{2}{c}{(12)}                   &
  \multicolumn{2}{c}{(13)}                   &
  \multicolumn{2}{c}{(14)}                   &
  \colhead{(15)}                             &
  \colhead{(16)}                             &
  \colhead{(17)}}

\startdata

WD0008$+$424 & 00 11 22.46 & $+$42 40 40.9 & A2-1 &  67 & 113 & 2008.57$-$2011.97 &  3.40 &   8 &  42.59 & 0.30 & 0.58 & 0.05 &  43.17 & 0.30 &  229.3 & 0.2 & 191.8 & 0.04 &  25.2 & EEV24 & \tablenotemark{a}  \\ 
WD0025$+$054 & 00 27 36.64 & $+$05 42 03.2 & A2-1 &  53 &  67 & 2012.62$-$2016.00 &  3.38 &  19 &  44.07 & 0.44 & 1.03 & 0.07 &  45.10 & 0.45 &  365.2 & 0.2 & 134.3 & 0.04 &  38.4 & EEV24 & \tablenotemark{a}  \\ 
WD0038$-$226 & 00 41 26.03 & $-$22 21 02.3 & A2-1 &  49 &  55 & 1992.79$-$1996.95 &  4.16 &   5 & 108.76 & 0.70 & 1.51 & 0.13 & 110.27 & 0.71 &  604.8 & 0.5 & 232.9 & 0.05 &  26.0 & Tek2K &   \\ 
WD0053$-$117 & 00 55 50.33 & $-$11 27 31.6 & A2-1 &  56 &  67 & 1992.79$-$1996.98 &  4.19 &   4 &  42.89 & 0.38 & 0.89 & 0.09 &  43.78 & 0.39 &  456.9 & 0.2 & 348.4 & 0.02 &  49.5 & Tek2K & \tablenotemark{a}  \\ 
WD0136$+$152 & 01 38 56.87 & $+$15 27 42.2 & A2-1 &  78 & 101 & 1992.67$-$1996.98 &  4.30 &  14 &  43.24 & 0.58 & 1.72 & 0.10 &  44.96 & 0.59 &   99.6 & 0.3 & 207.3 & 0.14 &  10.5 & Tek2K & \tablenotemark{a}  \\ 
WD0148$+$641 & 01 51 51.26 & $+$64 25 52.7 & A2-1 &  50 &  69 & 2012.61$-$2016.00 &  3.39 &  28 &  57.61 & 0.40 & 0.88 & 0.19 &  58.49 & 0.44 &  302.0 & 0.3 & 132.0 & 0.05 &  24.5 & EEV24 &   \\ 
WD0213$+$396 & 02 16 16.36 & $+$39 51 25.4 & ST-R &  35 &  70 & 1990.78$-$1995.82 &  5.03 &   5 &  49.92 & 0.82 & 0.83 & 0.07 &  50.75 & 0.82 &  186.0 & 0.4 & 240.4 & 0.14 &  17.4 & TI800 & \tablenotemark{a}  \\ 
WD0236$+$259 & 02 39 19.67 & $+$26 09 57.6 & A2-1 &  93 & 113 & 2008.72$-$2015.06 &  6.34 &  16 &  46.41 & 0.28 & 0.68 & 0.05 &  47.09 & 0.28 &  355.0 & 0.1 & 117.8 & 0.02 &  35.7 & EEV24 & \tablenotemark{a}  \\
\enddata

\tablenotetext{a}{New member of the 25 pc WD sample.}
\tablenotetext{b}{Object is discussed in Section
  \ref{subsec:comments}.}

\end{deluxetable}
\clearpage


\voffset0pt{}
\begin{deluxetable}{p{0.75in}r@{$\pm$}rccr@{$\pm$}rr@{$\pm$}rcccr@{$\pm$}rr@{$\pm$}rr@{$\pm$}rr@{$\pm$}rc}
\rotate 
\tabletypesize{\tiny} 
\tablecaption{Physical Parameters
\label{tab:parameters}}
\tablewidth{0pt}
\tablehead{
  \colhead{WD}                   & 
  \multicolumn{2}{c}{Adopted $\pi$\tablenotemark{a}}& 
  \colhead{No.}                  & 
  \colhead{}                     & 
  \multicolumn{2}{c}{$T_{\rm eff}$} & 
  \multicolumn{2}{c}{}           & 
  \colhead{Spectral}             & 
  \colhead{}                     & 
  \colhead{}                     & 
  \multicolumn{2}{c}{}           & 
  \multicolumn{2}{c}{}           & 
  \multicolumn{2}{c}{}           & 
  \multicolumn{2}{c}{Age\tablenotemark{b}} & 
  \colhead{}                    \\ 

  \colhead{Name}                 & 
  \multicolumn{2}{c}{(mas)}      & 
  \colhead{of $\pi$}          & 
  \colhead{Ref.}                 & 
  \multicolumn{2}{c}{(K)}        & 
  \multicolumn{2}{c}{\logg}      & 
  \colhead{Type}                 & 
  \colhead{Ref.}                 & 
  \colhead{Comp}                 & 
  \multicolumn{2}{c}{$M/$\msun}  & 
  \multicolumn{2}{c}{$M_V$}      & 
  \multicolumn{2}{c}{log $L/$\lsun} & 
  \multicolumn{2}{c}{(Gyr)}      & 
  \colhead{Notes}               \\ 

  \colhead{(1)}                  & 
  \multicolumn{2}{c}{(2)}        & 
  \colhead{(3)}                  & 
  \colhead{(4)}                  & 
  \multicolumn{2}{c}{(5)}        & 
  \multicolumn{2}{c}{(6)}        & 
  \colhead{(7)}                  & 
  \colhead{(8)}                  & 
  \colhead{(9)}                  & 
  \multicolumn{2}{c}{(10)}        & 
  \multicolumn{2}{c}{(11)}        & 
  \multicolumn{2}{c}{(12)}        & 
  \multicolumn{2}{c}{(13)}        & 
  \colhead{(14)}}                   

\startdata

WD2359$-$434  &  121.47  &  0.89 & 2 & 1,2         &   8390  &  170   & 8.37   & 0.02   &   DAP  &   16   & H          & 0.83   & 0.02   & 13.39 & 0.03 & $-$3.36  & 0.04   & 1.83   & 0.18   &     \\ 
WD0000$-$345  &   70.52  &  1.40 & 2 & 1,2         &   6280  &  100   & 8.18   & 0.03   &   DC   &   17   & He         & 0.68   & 0.03   & 14.27 & 0.05 & $-$3.76  & 0.04   & 3.23   & 0.35   &     \\ 
WD0008$+$424  &   43.17  &  0.30 & 1 & 3           &   7020  &  150   & 7.99   & 0.04   &   DA   &   18   & H          & 0.58   & 0.03   & 13.50 & 0.03 & $-$3.44  & 0.01   & 1.48   & 0.02   & \tablenotemark{c,\rm{d}}    \\ 
WD0011$-$721  &   52.77  &  1.10 & 1 & 1           &   6340  &  150   & 7.89   & 0.04   &   DA   &   19   & H          & 0.53   & 0.04   & 13.78 & 0.05 & $-$3.57  & 0.05   & 1.66   & 0.14   &     \\ 
WD0025$+$054  &   45.10  &  0.45 & 1 & 3           &   5650  &   60   & 7.97   & 0.02   &   DA   &   20   & H          & 0.57   & 0.02   & 14.42 & 0.04 & $-$3.81  & 0.02   & 2.61   & 0.17   &     \\ 
WD0034$-$602  &   41.41  &  1.57 & 1 & 1           &  14540  &  810   & 8.54   & 0.05   &   DA   &   21   & H          & 0.95   & 0.04   & 12.17 & 0.09 & $-$2.52  & 0.10   & 0.52   & 0.10   &     \\ 
WD0038$-$226  &  110.66  &  0.55 & 3 & 1,2,3       &   5230  &   60   & 7.92   & 0.03   &  DQpec &   22   & He($+$H)   & 0.52   & 0.03   & 14.71 & 0.03 & $-$3.94  & 0.03   & 4.23   & 0.34   & \tablenotemark{e}    \\ 
WD0046$+$051  &  232.37  &  1.18 & 3 & 1,2,4       &   6130  &  110   & 8.16   & 0.03   &   DZ   &   23   & He($+$H,Ca)& 0.67   & 0.02   & 14.21 & 0.03 & $-$3.79  & 0.04   & 3.45   & 0.36   & \tablenotemark{f}    \\
\enddata
\end{deluxetable}
\clearpage


\begin{deluxetable}{p{1.9in}r@{ $\pm$ }l}
\tabletypesize{\small}
\tablecaption{WD1242$-$105 Vital Parameters
\label{tab:wd1242_vitals}}
\tablewidth{0pt}
\tablehead{
%
  \multicolumn{3}{c}{Spectroscopic Results}}
\startdata

Period (days)\dotfill                   &  0.11885  &  0.00056   \\
$q$\dotfill                             &  0.685    &  0.013     \\
K1 (km s$^{\rm -1}$)\dotfill             &  184.6    &  2.3       \\
K2 (km s$^{\rm -1}$)\dotfill             &  126.5    &  1.9       \\
$\gamma$1 (km s$^{\rm -1}$)\dotfill      &  28.3     &  1.7       \\
$\gamma$2 (km s$^{\rm -1}$)\dotfill      &  44.8     &  1.9       \\
$\Delta\gamma$ (km s$^{\rm -1}$)\dotfill &  16.5     &  2.2     

\enddata
\end{deluxetable}


\begin{deluxetable}{p{3.0in}ccc}
\tabletypesize{\small}
\tablecaption{Contributions to the 25 pc WD Sample
\label{tab:contributions}}
\tablewidth{0pt}
\tablehead{
  \colhead{Parallax Program}         &
  \colhead{All $\pi$}                &
  \colhead{$\pi_{\rm err} \le$ 10\%} &
  \colhead{Refs.}} 

\startdata

Yale Parallax Catalog                &   104  & ~91  &  1       \\
{\it Hipparcos}                      &   ~~5  & ~~5  & 2,3,4,5  \\
Torino Observatory Parallax Program  &   ~~2  & ~~2  &  6       \\
Ducourant and Collaborators          &   ~~1  & ~~1  &  7       \\
\tableline
\\[-9pt]
~~~~Pre-2009 Sample & 112 & ~99  &               \\\\

Allegheny Observatory Parallax Program & ~~4 & ~~4  &  8  \\
\enddata

\tablenotetext{a}{As poor trigonometric parallax determinations, i.e.,
  parallax error $>$ 10\%, are updated with robust determinations, their
  counts are removed from this total.  At present, only four WD systems
  still have poorly constrained parallaxes.}

\tablerefs{
(1) \citealt{1995gcts.book.....V}, 
(2) \citealt{2007ASSL..350.....V},
(3) \citealt{2004ApJS..150..455G}, (4) \citealt{2005MNRAS.361L..15M}, 
(5) \citealt{2006A&A...456.1165C}, (6) \citealt{2003A&A...404..317S},
(7) \citealt{2007A&A...470..387D}, 
(8) \citealt{2009AJ....137..402G},
(9) \citealt{2009AJ....137.4109L},
(10) \citealt{2015MNRAS.449.3966G},
(11) \citealt{2016A&A...595A...4L},
(12) this work,
(13) \citealt{2009AJ....137.4547S}.}

\end{deluxetable}


\begin{deluxetable}{p{1.8in}cc}
\tabletypesize{\small}
\tablecaption{25 pc WD Sky Distribution
\label{tab:skydist}}
\tablewidth{0pt}
\tablehead{
  \multicolumn{1}{c}{Declination}    &
  \multicolumn{1}{c}{\# of}          &
  \multicolumn{1}{c}{\# of New}      \\

  \multicolumn{1}{c}{Range}          &
  \multicolumn{1}{c}{Systems}        &       
  \multicolumn{1}{c}{Systems}}       

\startdata

$+$90$^\circ$ to $+$30$^\circ$\dotfill        &       33   &  ~9  \\
$+$30$^\circ$ to $+$00$^\circ$\dotfill        &       37   &  11  \\
~~~Total (North)\dotfill      &\multicolumn{2}{c}{\bf 90}  \\[5pt]
$-$00$^\circ$ to $-$30$^\circ$\dotfill        &       28   &  13  \\
$-$30$^\circ$ to $-$90$^\circ$\dotfill        &       28   &  20  \\
~~~Total (South)\dotfill      &\multicolumn{2}{c}{\bf 89}  \\[3pt]

\enddata

\end{deluxetable}




\begin{thebibliography}{}

\bibitem[Alam et al.(2015)]{2015ApJS..219...12A} Alam, S., Albareti,
  F.~D., Allende Prieto, C., et al.\ 2015, \apjs, 219, 12

\bibitem[Aznar Cuadrado et al.(2004)]{2004A&A...423.1081A} Aznar
  Cuadrado, R., Jordan, S., Napiwotzki, R., et al.\ 2004, \aap, 423,
  1081

\bibitem[Bergeron et al.(1997)]{1997ApJS..108..339B} Bergeron, P.,
  Ruiz, M.~T., \& Leggett, S.~K.\ 1997, \apjs, 108, 339

\bibitem[Bergeron(2003)]{2003ApJ...586..201B} Bergeron, P.\ 2003,
  \apj, 586, 201

\bibitem[Bergeron et al.(2005)]{2005ApJ...625..838B} Bergeron, P.,
  Ruiz, M.~T., Hamuy, M., et al.\ 2005, \apj, 625, 838

\bibitem[Bergeron et al.(1995)]{1995ApJ...443..764B} Bergeron, P.,
  Saumon, D., \& Wesemael, F.\ 1995, \apj, 443, 764

\bibitem[Bertin(2011)]{2011ASPC..442..435B} Bertin, E.\ 2011,
  Astronomical Data Analysis Software and Systems XX, 442, 435

\bibitem[Bertin \& Arnouts(1996)]{1996A&AS..117..393B} Bertin, E., \&
  Arnouts, S.\ 1996, \aaps, 117, 393


\bibitem[Boyd et al.(2011)]{2011AJ....142...92B} Boyd, M.~R., Henry,
  T.~J., Jao, W.-C., Subasavage, J.~P., \& Hambly, N.~C.\ 2011, \aj,
  142, 92

\bibitem[Carpenter(2001)]{2001AJ....121.2851C} Carpenter, J.~M.\ 2001,
  \aj, 121, 2851

\bibitem[Casali et al.(2007)]{2007A&A...467..777C} Casali, M.,
  Adamson, A., Alves de Oliveira, C., et al.\ 2007, \aap, 467, 777

\bibitem[Chauvin et al.(2006)]{2006A&A...456.1165C} Chauvin, G.,
  Lagrange, A.-M., Udry, S., Fusco, T., Galland, F., Naef, D., Beuzit,
  J.-L., \& Mayor, M.\ 2006, \aap, 456, 1165

\bibitem[Dahn et al.(2002)]{2002AJ....124.1170D} Dahn, C.~C., Harris,
  H.~C., Vrba, F.~J., et al.\ 2002, \aj, 124, 1170

\bibitem[Debes \& Kilic(2010)]{2010AIPC.1273..488D} Debes, J.~H., \&
  Kilic, M.\ 2010, American Institute of Physics Conference Series,
  1273, 488

\bibitem[Debes et al.(2015)]{2015AJ....149..176D} Debes, J.~H., Kilic,
  M., Tremblay, P.-E., et al.\ 2015, \aj, 149, 176

\bibitem[Ducourant et al.(2007)]{2007A&A...470..387D} Ducourant, C.,
  Teixeira, R., Hambly, N.~C., Oppenheimer, B.~R., Hawkins, M.~R.~S.,
  Rapaport, M., Modolo, J., \& Lecampion, J.~F.\ 2007, \aap, 470, 387

\bibitem[Dufour et al.(2005)]{2005ApJ...627..404D} Dufour, P.,
  Bergeron, P., \& Fontaine, G.\ 2005, \apj, 627, 404

\bibitem[Dufour et al.(2007)]{2007ApJ...663.1291D} Dufour, P.,
  Bergeron, P., Liebert, J., et al.\ 2007, \apj, 663, 1291

\bibitem[Farihi et al.(2011)]{2011MNRAS.413.2559F} Farihi, J., Dufour,
  P., Napiwotzki, R., \& Koester, D.\ 2011, \mnras, 413, 2559

\bibitem[Farihi et al.(2009)]{2009ApJ...694..805F}Farihi, J., Jura,
  M., \& Zuckerman, B.\ 2009, \apj, 694, 805

\bibitem[Fontaine et al.(2001)]{fon01} Fontaine, G., Brassard, P., \&
  Bergeron, P. 2001, \pasp, 113, 409

\bibitem[Gaia Collaboration et al.(2016a)]{2016A&A...595A...2G} Gaia
  Collaboration, Brown, A.~G.~A., Vallenari, A., et al.\ 2016a, \aap,
  595, A2

\bibitem[Gaia Collaboration et al.(2016b)]{2016A&A...595A...1G} Gaia
  Collaboration, Prusti, T., de Bruijne, J.~H.~J., et al.\ 2016b, \aap,
  595, A1


\bibitem[Gatewood \& Coban(2009)]{2009AJ....137..402G} Gatewood, G.,
  \& Coban, L.\ 2009, \aj, 137, 402

\bibitem[Giammichele et al.(2012)]{2012ApJS..199...29G} Giammichele,
  N., Bergeron, P., \& Dufour, P.\ 2012, \apjs, 199, 29

\bibitem[Gianninas et al.(2011)]{2011ApJ...743..138G} Gianninas, A.,
  Bergeron, P., \& Ruiz, M.~T.\ 2011, \apj, 743, 138

\bibitem[Gianninas et al.(2015)]{2015MNRAS.449.3966G} Gianninas, A.,
  Curd, B., Thorstensen, J.~R., et al.\ 2015, \mnras, 449, 3966

\bibitem[Gould \& Chanam{\'e}(2004)]{2004ApJS..150..455G} Gould, A.,
  \& Chanam{\'e}, J.\ 2004, \apjs, 150, 455

\bibitem[Graham(1982)]{1982PASP...94..244G} Graham, J.~A.\ 1982,
  \pasp, 94, 244

\bibitem[Hambly et al.(2008)]{2008MNRAS.384..637H} Hambly, N.~C.,
  Collins, R.~S., Cross, N.~J.~G., et al.\ 2008, \mnras, 384, 637

\bibitem[Hansen(1998)]{1998Natur.394..860H} Hansen, B.~M.~S.\ 1998,
  \nat, 394, 860

\bibitem[Harris et al.(2016)]{2016AJ....152..118H} Harris, H.~C.,
  Dahn, C.~C., Zacharias, N., et al.\ 2016, \aj, 152, 118


\bibitem[Henry et al.(2004)]{2004AJ....128.2460H} Henry, T.~J.,
  Subasavage, J.~P., Brown, M.~A., Beaulieu, T.~D., Jao, W., \&
  Hambly, N.~C.\ 2004, \aj, 128, 2460

\bibitem[Hewett et al.(2006)]{2006MNRAS.367..454H} Hewett, P.~C.,
  Warren, S.~J., Leggett, S.~K., \& Hodgkin, S.~T.\ 2006, \mnras, 367,
  454

\bibitem[Hodgkin et al.(2009)]{2009MNRAS.394..675H} Hodgkin, S.~T.,
  Irwin, M.~J., Hewett, P.~C., \& Warren, S.~J.\ 2009, \mnras, 394,
  675

\bibitem[Holberg et al.(2006)]{holberg06} Holberg, J. B., \& Bergeron,
  P. 2006, \aj, 132, 1223

\bibitem[Holberg et al.(2016)]{2016MNRAS.462.2295H} Holberg, J.~B.,
  Oswalt, T.~D., Sion, E.~M., \& McCook, G.~P.\ 2016, \mnras, 462,
  2295

\bibitem[Jao et al.(2005)]{2005AJ....129.1954J} Jao, W.-C., Henry,
  T.~J., Subasavage, J.~P., Brown, M.~A., Ianna, P.~A., Bartlett,
  J.~L., Costa, E., \& M{\'e}ndez, R.~A.\ 2005, \aj, 129, 1954


\bibitem[Jefferys et al.(1988)]{1988CeMec..41...39J} Jefferys, W.~H.,
  Fitzpatrick, M.~J., \& McArthur, B.~E.\ 1988, Celestial Mechanics,
  41, 39

\bibitem[Kawka \& Vennes(2006)]{2006ApJ...643..402K} Kawka, A., \&
  Vennes, S.\ 2006, \apj, 643, 402

\bibitem[Kawka \& Vennes(2012)]{2012MNRAS.425.1394K} Kawka, A., \&
  Vennes, S.\ 2012, \mnras, 425, 1394

\bibitem[Kawka et al.(2007)]{2007ApJ...654..499K} Kawka, A., Vennes,
  S., Schmidt, G.~D., Wickramasinghe, D.~T., \& Koch, R.\ 2007, \apj,
  654, 499

\bibitem[Kawka et al.(2004)]{2004AJ....127.1702K} Kawka, A., Vennes,
  S., \& Thorstensen, J.~R.\ 2004, \aj, 127, 1702

\bibitem[Kilic et al.(2009)]{2009ApJ...696.2094K} Kilic, M., Kowalski,
  P.~M., Reach, W.~T., \& von Hippel, T.\ 2009, \apj, 696, 2094

\bibitem[Kilkenny et al.(1988)]{1988SAAOC..12....1K} Kilkenny, D.,
  Heber, U., \& Drilling, J.~S.\ 1988, South African Astronomical
  Observatory Circular, 12,

\bibitem[Kirkpatrick et al.(1995)]{1995AJ....109..797K} Kirkpatrick,
  J.~D., Henry, T.~J., \& Simons, D.~A.\ 1995, \aj, 109, 797

\bibitem[Koerner et al.(2003)]{2003AAS...203.4207K} Koerner, D.~W.,
  Henry, T.~J., Fuhrman, L.~A., et al.\ 2003, Bulletin of the American
  Astronomical Society, 35, 42.07

\bibitem[Koester et al.(1997)]{1997A&A...320L..57K} Koester, D.,
  Provencal, J., \& Shipman, H.~L.\ 1997, \aap, 320, L57

\bibitem[Koester et al.(2005)]{2005A&A...432.1025K} Koester, D.,
  Rollenhagen, K., Napiwotzki, R., et al.\ 2005, \aap, 432, 1025

\bibitem[Koester et al.(2011)]{2011A&A...530A.114K} Koester, D.,
  Girven, J., G{\"a}nsicke, B.~T., \& Dufour, P.\ 2011, \aap, 530,
  A114

\bibitem[Landolt(1992)]{1992AJ....104..340L} Landolt, A.~U.\ 1992,
  \aj, 104, 340
 
\bibitem[Landolt(2007)]{2007AJ....133.2502L} Landolt, A.~U.\ 2007,
  \aj, 133, 2502

\bibitem[Landolt(2013)]{2013AJ....146..131L} Landolt, A.~U.\ 2013,
  \aj, 146, 131

\bibitem[Lawrence et al.(2007)]{2007MNRAS.379.1599L} Lawrence, A.,
  Warren, S.~J., Almaini, O., et al.\ 2007, \mnras, 379, 1599

\bibitem[L{\'e}pine et al.(2003)]{2003AJ....125.1598L} L{\'e}pine, S.,
  Rich, R.~M., \& Shara, M.~M.\ 2003, \aj, 125, 1598

\bibitem[L{\'e}pine et al.(2005)]{2005ApJ...633L.121L} L{\'e}pine, S.,
  Rich, R.~M., \& Shara, M.~M.\ 2005, \apjl, 633, L121

\bibitem[L{\'e}pine \& Shara(2005)]{2005AAS...20715001L} L{\'e}pine,
  S., \& Shara, M.~M.\ 2005, Bulletin of the American Astronomical
  Society, 37, 150.01

\bibitem[L{\'e}pine et al.(2002)]{2002AJ....124.1190L} L{\'e}pine, S.,
  Shara, M.~M., \& Rich, R.~M.\ 2002, \aj, 124, 1190

\bibitem[L{\'e}pine et al.(2009)]{2009AJ....137.4109L} L{\'e}pine, S.,
  Thorstensen, J.~R., Shara, M.~M., \& Rich, R.~M.\ 2009, \aj, 137,
  4109

\bibitem[Liebert \& Strittmatter(1977)]{1977ApJ...217L..59L} Liebert,
  J., \& Strittmatter, P.~A.\ 1977, \apjl, 217, L59

\bibitem[Limoges et al.(2015)]{2015ApJS..219...19L} Limoges, M.-M.,
  Bergeron, P., \& L{\'e}pine, S.\ 2015, \apjs, 219, 19

\bibitem[Lindegren et al.(2016)]{2016A&A...595A...4L} Lindegren, L.,
  Lammers, U., Bastian, U., et al.\ 2016, \aap, 595, A4

\bibitem[Luyten(1972)]{1972PMMin..30....1L} Luyten, W.~J.\ 1972,
  Proper Motion Survey, University of Minnesota, 30,

\bibitem[Luyten \& La Bonte(1972)]{1972PMMin..32....1L} Luyten, W.~J.,
  \& La Bonte, A.~E.\ 1972, Proper Motion Survey, University of
  Minnesota, 32,

\bibitem[Luyten(1974)]{1974PMMin..37....1L} Luyten, W.~J.\ 1974,
  Proper Motion Survey, University of Minnesota, 37,

\bibitem[Luyten(1979a)]{1979nlcs.book.....L} Luyten, W.~J.\ 1979a, New
  Luyten catalogue of stars with proper motions larger than two tenths
  of an arcsecond; and first supplement; NLTT.~(Minneapolis (1979));
  Label 12 = short description; Label 13 = documentation by Warren;
  Label 14 = catalogue, Strasbourg version,

\bibitem[Luyten(1979b)]{1979lccs.book.....L} Luyten, W.~J.\ 1979b,
  Minneapolis: University of Minnesota, 1979, 2nd ed., LHS

\bibitem[McGraw \& Robinson(1975)]{1975ApJ...200L..89M} McGraw, J.~T.,
  \& Robinson, E.~L.\ 1975, \apjl, 200, L89

\bibitem[Monet \& Dahn(1983)]{1983AJ.....88.1489M} Monet, D.~G., \&
  Dahn, C.~C.\ 1983, \aj, 88, 1489

\bibitem[Monet et al.(1992)]{1992AJ....103..638M} Monet, D.~G., Dahn,
  C.~C., Vrba, F.~J., et al.\ 1992, \aj, 103, 638

\bibitem[Mugrauer \& Neuh{\"a}user(2005)]{2005MNRAS.361L..15M}
  Mugrauer, M., \& Neuh{\"a}user, R.\ 2005, \mnras, 361, L15

\bibitem[Oppenheimer et al.(2001)]{2001Sci...292..698O} Oppenheimer,
  B.~R., Hambly, N.~C., Digby, A.~P., Hodgkin, S.~T., \& Saumon,
  D.\ 2001, Science, 292, 698


\bibitem[Press et al.(1992)]{pressetal92} Press, W. H., Teukolsky,
  S. A., Vetterling, W. T., \& Flannery, B. P. 1992, Numerical Recipes
  in FORTRAN, 2nd edition (Cambridge: Cambridge University Press), 644


\bibitem[Probst et al.(2004)]{2004SPIE.5492.1716P} Probst, R.~G.,
  Gaughan, N., Abraham, M., et al.\ 2004, \procspie, 5492, 1716

\bibitem[Reimers et al.(1996)]{1996A&A...311..572R} Reimers, D.,
  Jordan, S., Koester, D., et al.\ 1996, \aap, 311, 572

\bibitem[Ruiz \& Bergeron(2001)]{2001ApJ...558..761R} Ruiz, M.~T., \&
  Bergeron, P.\ 2001, \apj, 558, 761

\bibitem[Ruiz et al.(2001)]{2001ApJS..133..119R} Ruiz, M.~T.,
  Wischnjewsky, M., Rojo, P.~M., \& Gonzalez, L.~E.\ 2001, \apjs, 133,
  119

\bibitem[Salim et al.(2004)]{2004ApJ...601.1075S} Salim, S., Rich,
  R.~M., Hansen, B.~M., et al.\ 2004, \apj, 601, 1075

\bibitem[Saumon et al.(1994)]{1994ApJ...424..333S} Saumon, D.,
  Bergeron, P., Lunine, J.~I., Hubbard, W.~B., \& Burrows, A.\ 1994,
  \apj, 424, 333

\bibitem[Saumon \& Jacobson(1999)]{1999ApJ...511L.107S} Saumon, D., \&
  Jacobson, S.~B.\ 1999, \apjl, 511, L107

\bibitem[Sayres et al.(2012)]{2012AJ....143..103S} Sayres, C.,
  Subasavage, J.~P., Bergeron, P., et al.\ 2012, \aj, 143, 103

\bibitem[Schlafly \& Finkbeiner(2011)]{2011ApJ...737..103S} Schlafly,
  E.~F., \& Finkbeiner, D.~P.\ 2011, \apj, 737, 103

\bibitem[Schmidt et al.(2001)]{2001MNRAS.328..203S} Schmidt, G.~D.,
  Vennes, S., Wickramasinghe, D.~T., \& Ferrario, L.\ 2001, \mnras,
  328, 203

\bibitem[Scholz et al.(2002)]{2002ApJ...565..539S} Scholz, R.-D.,
  Szokoly, G.~P., Andersen, M., Ibata, R., \& Irwin, M.~J.\ 2002,
  \apj, 565, 539

\bibitem[Silvestri et al.(2002)]{2002AJ....124.1118S} Silvestri, N.~M., 
Oswalt, T.~D., \& Hawley, S.~L.\ 2002, \aj, 124, 1118 

\bibitem[Silvotti et al.(2015)]{2015ASPC..493..455S} Silvotti, R.,
  Sozzetti, A., Lattanzi, M., \& Morbidelli, R.\ 2015, 19th European
  Workshop on White Dwarfs, 493, 455

\bibitem[Sion et al.(1983)]{1983ApJ...269..253S} Sion, E.~M.,
  Greenstein, J.~L., Landstreet, J.~D., et al.\ 1983, \apj, 269, 253

\bibitem[Shulov \& Kopatskaya(1974)]{1974AFZ...10..117S} Shulov,
  O.~S., \& Kopatskaya, E.~N.\ 1974, {\it Astrofizika}, 10, 117

\bibitem[Smart et al.(2003)]{2003A&A...404..317S} Smart, R.~L., et
  al.\ 2003, \aap, 404, 317

\bibitem[Stetson(1987)]{1987PASP...99..191S} Stetson, P.~B.\ 1987,
  \pasp, 99, 191

\bibitem[Strand(1964)]{1964S&T....27..204S} Strand, K.~A.\ 1964,
  \skytel, 27,

\bibitem[Subasavage et al.(2007)]{2007AJ....134..252S} Subasavage,
  J.~P., Henry, T.~J., Bergeron, P., et al.\ 2007, \aj, 134, 252

\bibitem[Subasavage et al.(2008)]{2008AJ....136..899S} Subasavage,
  J.~P., Henry, T.~J., Bergeron, P., Dufour, P., \& Hambly,
  N.~C.\ 2008, \aj, 136, 899

\bibitem[Subasavage et al.(2009)]{2009AJ....137.4547S} Subasavage,
  J.~P., Jao, W.-C., Henry, T.~J., et al.\ 2009, \aj, 137, 4547

\bibitem[Tremblay \& Bergeron(2009)]{2009ApJ...696.1755T} Tremblay,
  P.-E., \& Bergeron, P.\ 2009, \apj, 696, 1755

\bibitem[Vennes \& Kawka(2003)]{2003ApJ...586L..95V} Vennes, S., \&
  Kawka, A.\ 2003, \apjl, 586, L95

\bibitem[van Altena et al.(1995)]{1995gcts.book.....V} van Altena,
  W.~F., Lee, J.~T., \& Hoffleit, E.~D.\ 1995, New Haven, CT: Yale
  University Observatory, |c1995, 4th ed., completely revised and
  enlarged

\bibitem[van Leeuwen(2007)]{2007ASSL..350.....V} van Leeuwen,
  F.\ 2007, Hipparcos, the New Reduction of the Raw Data.~By Floor van
  Leeuwen, Institute of Astronomy, Cambridge University, Cambridge, UK
  Series: Astrophysics and Space Science Library, Vol.~ 350 20
  Springer Dordrecht

\bibitem[Xu et al.(2014)]{2014ApJ...783...79X} Xu, S., Jura, M.,
  Koester, D., Klein, B., \& Zuckerman, B.\ 2014, \apj, 783, 79

\bibitem[Zuckerman \& Becklin(1987)]{1987Natur.330..138Z} Zuckerman,
  B., \& Becklin, E.~E.\ 1987, \nat, 330, 138

\bibitem[Zuckerman et al.(2003)]{2003ApJ...596..477Z} Zuckerman, B.,
  Koester, D., Reid, I.~N., H{\"u}nsch, M.\ 2003, \apj, 596, 477

%
%
%
%
%
%
%
%
%
%
%
%
%
%
%
%
%
%
%

%
%
%
%
%
%
%
%

%
%
%
%
%
%
%
%
%
%
%
%
%
%
%
%
%

\end{thebibliography}
\end{document}